\documentclass[a4paper,preprintnumbers,floatfix,superscriptaddress,aps,pra,twocolumn,showpacs,notitlepage,longbibliography
]{revtex4-2}

\usepackage[utf8]{inputenc}
\usepackage[english]{babel}
\usepackage[a4paper,margin=2.0cm]{geometry}


\usepackage{indentfirst}
\usepackage{graphicx}
\usepackage{booktabs}
\usepackage{float}
\usepackage{siunitx}
\usepackage{lipsum}
\usepackage{listings}
\usepackage{amsmath}
\usepackage{amsfonts}
\usepackage{tasks}
\usepackage{amssymb}
\usepackage{wasysym}
\usepackage[T1]{fontenc}
\date{}
\usepackage{tikz}
\usepackage{comment}
\usepackage{mathtools}
\usepackage{xcolor}
\usepackage{hyperref}
\usepackage{url}
\usepackage{natbib}

\newcommand{\ie}{\textit{i.e.}}

\newcommand{\mean}[1]{\left<#1 \right>}

\newcommand{\ba}[1]{\begin{align} #1 \end{align}}

\begin{document}

\title{Information causality in multipartite scenarios}
\author{Lucas Pollyceno}
\email{lpolly@ifi.unicamp.br}
\affiliation{Instituto de Física “Gleb Wataghin”, Universidade Estadual de Campinas, CEP 13083-859, Campinas, Brazil}
\author{Rafael Chaves}
\affiliation{International Institute of Physics, Federal University of Rio Grande do Norte, 59070-405 Natal, Brazil}
\affiliation{School of Science and Technology, Federal University of Rio Grande do Norte, 59078-970 Natal, Brazil}
\author{Rafael Rabelo}
\affiliation{Instituto de Física “Gleb Wataghin”, Universidade Estadual de Campinas, CEP 13083-859, Campinas, Brazil}

\date{\today}

\begin{abstract}
Bell nonlocality is one of the most intriguing and counter-intuitive phenomena displayed by quantum systems. Interestingly, such stronger-than-classical quantum correlations are somehow constrained, and one important question to the foundations of quantum theory is whether there is a physical, operational principle responsible for those constraints. One candidate is the information causality principle, which, in some particular cases, is proven to hold for quantum systems and to be violated by stronger-than-quantum correlations. In multipartite scenarios, though, it is known that the original formulation of the information causality principle fails to detect even extremal stronger-than-quantum correlations, thus suggesting that a genuinely multipartite formulation of the principle is necessary. In this work, we advance towards this goal, reporting a new formulation of the information causality principle in multipartite scenarios. By proposing a change of perspective, we obtain multipartite informational inequalities that work as necessary criteria for the principle to hold. We prove that such inequalities hold for all quantum resources, and forbid some stronger-than-quantum ones. Finally, we show that our approach can be strengthened if multiple copies of the resource are available, or, counter-intuitively, if noisy communication channels are employed.
\end{abstract}

\maketitle

%\begin{multicols}{2}

\section{Introduction}
% Quantum has mathematical axioms, and nonlocality
It is undeniable that quantum theory is one of the most successful scientific theories ever developed. Its mathematical formalism and axioms, although counter-intuitive, have led to extremely precise and, oftentimes, intriguing predictions, which have been confirmed experiment after experiment. Among the most counter-intuitive phenomena that quantum systems can display, Bell nonlocality \cite{Bell1964, Brunner_etal2014} is one of the most fascinating.

Bell nonlocality refers to stronger-than-classical correlations between outcomes of measurements performed on space-like separated systems. However, as noted by Tsirelson \cite{Tsirelson_lim}, and, later, by Popescu and Rohrlich \cite{PRbox}, nonlocal correlations displayed by quantum systems are limited and are not as strong as they could be, in principle.  From a formal viewpoint, such limitations follow from the mathematical axioms of quantum theory. From a physical viewpoint, though, it would be of interest to the foundations of quantum theory to identify operational axioms that could explain the limits of quantum nonlocality and, ultimately, lead to the derivation of the mathematical axioms of the theory from first principles.

% Operational axioms have been proposed
In the last couple of decades, several principles have been proposed with the goal to explain why quantum theory is not more nonlocal. Among them, are the principle of nontrivial communication complexity \cite{NontrivialCC}, the principle of macroscopic locality \cite{ML}, and the principle of local orthogonality \cite{OL}. Although all of the cited candidate principles are very good at identifying and forbidding unreasonable consequences of stronger-than-quantum correlations, they are provably not capable of excluding \emph{all} stronger-than-quantum correlations \cite{navascues2015almost}. 

% Information causality is an interesting one
One candidate principle that may be capable of singling out the nonlocal correlations allowed by quantum theory from more nonlocal ones is the information causality (IC) principle \cite{IC}. Roughly speaking, the principle states that in a communication scenario, the receiver's available information concerning the sender's initial set of data can not exceed the message information amount. The violation of the principle would allow, for instance, one to receive an amount of information corresponding to one page of a book, and choose, afterwards, which page the receiver would like to read.

It is well known that all quantum correlations, despite nonlocal, obey the information causality principle. It is also known that \emph{some} stronger-than-quantum correlations violate the principle \cite{IC}. It is unclear, though, if \emph{all} stronger-than-quantum correlations violate it. The main reason for this is the fact that the principle, although relatively intuitive, is very hard to formalize in the form of a mathematical criteria. A sufficient criterion for the violation of the principle was proposed in Ref.~\cite{IC}, but it has been later proved that there are non-quantum correlations that do not violate it \cite{Allcock_2009}. Since then, other techniques have been developed to generate stronger criteria for the information causality principle \cite{Chaves_2015, yu2022information}; but, so far, none of them has shown to be strong enough to characterize the exact set of quantum correlations, even in simple scenarios.

% However, an operational axiom has to be multipartite
Another requirement that has been discovered recently is that any operational principle has to be genuinely multipartite to correctly retrieve all quantum nonlocal correlations \cite{PhysRevLett.107.210403}. This observation complements the finding that the bipartite formulation of the original IC proposal cannot be applied to exclude even some extremal tripartite stronger-than-quantum correlations \cite{Yang_2012}. Hence, it is clear that a genuinely multipartite formulation of information causality is necessary for it to be a valid and tight operational principle for quantum theory.

% In this work...
In this work, we propose a novel multipartite perspective for the information causality principle. First, we establish a multipartite communication task that, in a sense, generalizes the random access code (RAC) task associated with the original IC formulation of Ref.~\cite{IC}. Thus, we present new multipartite informational-theoretic criteria, which ensure IC in the new proposed scenario. We then prove that the inequalities hold for all quantum nonlocal resources, and are violated by some stronger-than-quantum ones. Also, if many copies of the nonlocal resource are available, we show that by applying the concatenation approach presented in Ref.~\cite{IC}, and in analogy to the result obtained in such reference, one obtains stronger criteria for IC. In addition, we show that our findings are in agreement with the recent results reported in Ref.~\cite{ICnoisy}, where it is shown that the employment of noisy communication channels leads to the same constraints as the concatenation procedure mentioned.

\section{Bipartite Information causality}

The \textit{information causality} scenario \cite{IC}, considers a strictly bipartite communication task: Alice encodes a bit-string $\mathbf{x}$ of length $n$ in a message $M$ of $m$ bits (where $m<n$), and sends this message to Bob; he, then, decodes the message and produces a guess $G_i$ about a randomly selected bit $X_i$ (the i-th bit of the string $\mathbf{x}$) of Alice. Within this context, the information causality principle states that \textit{Bob's information gain about the initial $n$ bits of Alice, considering all their possible local as well as pre-established shared resources, cannot be greater than the number of bits $m$, sent by Alice.} The principle, proven to hold quantum mechanically, is mathematically captured by an information inequality, written in terms of Shannon entropies, as
\ba{\label{firstIC}\sum_{i=1}^{n} I(X_i:G_i) \le H(M),}
where $H(M)=-\sum_{m}p(m)\log_2{p(m)}$ is the Shannon entropy of the random variable $M$ described by the probability distribution $p(m)=p(M=m)$. In turn $I(X_i:G_i)=H(X_i)+H(G_i)-H(X_i,G_i)$ stands for the mutual information between Alice's bit $X_i$ and Bob's guess $G_i$.

The information causality inequality \eqref{firstIC} can be violated by post-quantum correlations, that is, correlations incompatible with the quantum mechanical rules. A paradigmatic example is that of a Popescu-Rohrlich(PR)-box \cite{PRbox}, a non-signalling (NS) correlation described by the probability distribution
    \begin{equation}\label{PRbox}
        p(a,b|x,y) = \left\{\begin{aligned}
        1/2 & \quad \text{if} \quad a\oplus b  = xy ; \\
        0& \quad\text{else}.
        \end{aligned}\right.
    \end{equation}
To verify that is indeed the case, it is sufficient to consider Alice has two input bits, described by the random variables $X_1$ and $X_2$, and consider the following protocol in the IC scenario. Alice inputs $x_1 \oplus x_2$ on her share of the PR-box, obtaining outcome $a$ that is then encoded in the message $m=a\oplus x_1$ to Bob. Bob inputs $y=0$ if his aim is to guess the bit $x_1$ and $y=1$ if wants to guess the $x_2$ of Alice, using as his guess $g_i=m \oplus b$. Using the definition of the PR-box, if $y=0$ we see that $g_1= x_1 \oplus a \oplus b= x_1$. If $y=1$, $g_2= x_1 \oplus a \oplus b= x_1 \oplus x_1 \oplus x_2=x_2$, implying that $I(X_1:G_1)=I(X_2:G_2)=H(M)=1$, thus violating the IC inequality \eqref{firstIC}.

Notice that the simple protocol above employs a single copy of the distribution $p(a,b\vert x,y)$, which operationally can be understood as a black-box taking local inputs and producing correlated outputs.
By introducing a concatenation procedure (a version of which will be detailed below), one can also consider the case of multiple copies of identical binary-input/binary-output non-signaling boxes. In particular, the  concatenation procedure introduced in \cite{IC} shows that the IC inequality implies another constraint for non-signaling correlations, given by 
\ba{\label{Uffink}
E_I^2 + E_{II}^2 \le 1,}
where $E_j= 2 P_j -1$ is defined in terms of the conditional probabilities $p(a,b|x,y)$ as
\begin{subequations}\ba{
P_I &= \frac{1}{2}[p(a\oplus b = 0|0,0) + p(a\oplus b = 0|1,0)];\\
P_{II} &= \frac{1}{2}[p(a\oplus b = 0|0,1) + p(a\oplus b = 1|1,1)].}\end{subequations}
In fact, this constraint is equivalent to the bipartite quadratic Bell inequality, the so-called Uffink's inequality \cite{UffinkP} (interestingly, however, in the next sections we will show that for multipartite scenarios such equivalence no longer holds). 

Of particular relevance is the fact that this mapping from the IC inequality \eqref{firstIC} to Uffink's inequality \eqref{Uffink}, proves that any correlation beyond the Tsirelson's limit for the Clauser-Horne-Shimony-Holtz (CHSH) inequality \cite{CHSH} will violate the information causality principle and thus witness its incompatibility with quantum theory. More precisely, as proven by Tsirelson \cite{cirel1980quantum}, the classically valid CHSH inequality 
\begin{equation}
    \mathrm{CHSH}= \mean{A_0B_0}+\mean{A_0B_1}+\mean{A_1B_0}-\mean{A_1B_1} \leq 2,
\end{equation}
achieves a maximum value in quantum theory of $\mathrm{CHSH}_{Q} = 2\sqrt{2}$. A PR-box in turn leads to $\mathrm{CHSH}_{NS}=4$. A direct analysis shows that any distribution achieving $\mathrm{CHSH} > \mathrm{CHSH}_{Q}$ violates Uffink's inequality and thus has its post-quantum nature witnessed by the information causality principle. One should remark, however, that it remains unclear whether the whole set of quantum correlations can be recovered from the IC principle \cite{Allcock_2009,navascues2015almost,brito2019nonlocality}. Interestingly, there are known post-quantum correlations known to not violate Uffink's inequality \cite{Chaves_2015,rai2019geometry}.

Motivated by that insufficiency of the standard formulation of the IC principle, a general informational-geometric approach has been introduced in \cite{Chaves_2015} and shown to lead to stronger IC information inequalities. For example, the inequality given by
%\begin{widetext}
\begin{equation}
\begin{aligned}\label{recentIC} \sum_{i=1}^{n}I(X_i : G_i,M) + \sum_{i=2}^{n}I(X_1 : X_i| G_i,M) \\ \le H(M) + \sum_{i=2}^{n}H(X_i) - H(X_0,...,X_{n-1}).\end{aligned}\end{equation}
%\end{widetext}
The original IC inequality \eqref{firstIC} is a particular case of \eqref{recentIC}. As a matter of fact, the stronger IC inequality \eqref{recentIC} was proven, in some cases, to be even stronger than Uffink's inequality \eqref{Uffink}. More precisely, with a single copy, inequality \eqref{recentIC} can detect the post-quantumness of correlations that cannot be detected by \eqref{Uffink} even in the asymptotic regime of infinite copies of the correlation under test.
\begin{figure}[t!]
    \centering
    \scalebox{0.8}{
    \begin{tikzpicture}{}
      \node[circle,draw, very thick](x0) at (0,6) {\Large{$X_1^1$}};
      \node[very thick] at (0,5.1) {\Large{$\vdots$}};
      \node[circle,draw, very thick](x1) at (0,4) {\Large{$X_n^1$}};
      \node[circle,draw, very thick](mx) at (2.5,5.25) {\Large{$M_1$}};
      \node[very thick] at (2.5,2.75) {\Huge{$\vdots$}};

      \node[circle,draw, very thick](y0) at (0,1.5) {\large{$X_1^{N-1}$}};
      \node[very thick] at (0,0.6) {\Large{$\vdots$}};
      \node[circle,draw, very thick](y1) at (0,-0.5) {\large{$X_n^{N-1}$}};
      \node[circle,draw, very thick](my) at (2.5,0.25) {\large{$M_{N-1}$}};

      \node[draw,inner sep=10pt](l) at (4.5,2.75) {\Large{$\rho$}};

      \node[circle,draw, very thick](g0) at (7.5,2.75) {\Large{$G_j$}};
      \node[circle,draw, very thick](j) at (7.5,0.75) {\Large{$j$}};
      \node(z) at (7,2.75) {};

      \draw[very thick, ->] (x0) -- (mx);
      \draw[very thick, ->] (x1) -- (mx);
      \draw[very thick, ->] (mx) to[out=0,in=135] (z);
      \draw[very thick, ->] (y0) -- (my);
      \draw[very thick, ->] (y1) -- (my);
      \draw[very thick, ->] (my) to[out=0,in=225] (z);
      \draw[very thick, ->] (l) -- (mx);
      \draw[very thick, ->] (l) -- (my);
      \draw[very thick, ->] (l) -- (z);
      \draw[very thick, ->] (j) -- (g0);
      \end{tikzpicture}}
    \caption{Quantum causal structure, described as a DAG, associated with the multipartite information causality scenario.
    }
    \label{NewScenarioIC}
\end{figure}
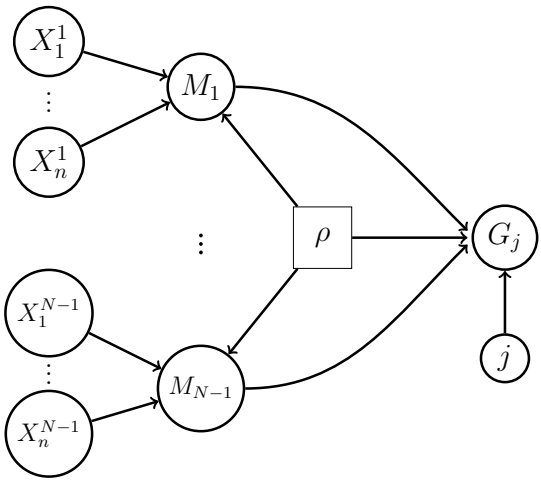

Building upon the original IC criterion, a more recent approach \cite{ICnoisy}, generalized it to consider noisy channels between Alice and Bob, namely
\begin{equation}
\sum_{i=1}^{n} I(X_i:G_i) \le C,
\label{firstnoisy}    
\end{equation}
where $C\equiv I(M:M')$ is the noisy channel capacity defined in terms of Shannon's mutual information, and $M'$ is the message that reaches the receiver after passing through the channel. Contrary to the original formulation, this new approach proposes to search for the strongest non-signaling correlations allowed by IC for every possible noisy channel between the parts. A generalization that allows recovering standard results, for instance,  the Tsirelson’s bound implied by Uffink’s inequality \eqref{Uffink}, most importantly, however, without the need of a concatenation procedure, that is, in the single copy level. This stems from the fact that as the channel's noise is increased, the bound in \eqref{firstnoisy} becomes stronger, capable of detecting the post-quantum nature of correlations that cannot be witnessed in the noise-free version. 

The main goal of this paper, as will be detailed in the following, is to generalize all such results, stated so far only for the bipartite scenario, into the multipartite setting

\section{New criteria for multipartite information causality}\label{newIC}
 
Our first goal is to introduce a natural generalization of the bipartite IC inequality to the multipartite scenario. For that, we will closely follow the information-theoretic approach based on quantum causal structures envisioned in \cite{Chaves_2015}. In this approach, information principles such as information causality are nothing else than entropic constraints arising from imposing a quantum description to a given causal structure. As such, each quantum causal structure will have associated with it a given set of entropic inequalities, each of which can be interpreted as an information-theoretical principle.

In this work, we consider a particular class of quantum causal structures that naturally generalize the known bipartite scenario. Consider $N$ parts, among which $N-1$ are senders in possession of their respective bit-strings $\mathbf{x}^k = (X_1^k, X_2^k, \cdots, X_n^k)$, where $k\in \{1, 2, \cdots, N-1\}$. Each sender encodes a classical message $M_k$ of size $m < n$ to the $N^{th}$-part, the receiver who has to compute one out of $n$ possible bits functions $f_j (X_j^1, X_j^2, \cdots, X_j^{N-1})$, by producing the guess $G_j$, where $j\in \{1, \cdots, n\}$. This scenario is illustrated for the tripartite case, as a directed acyclic graph (DAG), in Fig. \ref{NewScenarioIC}. As proven in the Appendix \ref{truthfulness}, considering that additionally to any local operations, every part may explore their pre-established correlations mediated by a joint quantum state $\rho$, the following multipartite version of information causality holds:
\begin{widetext}
\begin{align}\label{ICmulti}
    \sum_k^{N-1} \sum_i^n I(X_i^k : X_i^1, \dots,X_i^{k-1}, X_i^{k+1}, \dots, X_i^{N-1}, G_i)   \le  
    H(M_1,\dots,M_{N-1})+ \sum_k^{N-1} \sum_i^n I(X_{i+1}^k, \dots, X_n^k :  X_i^k).
\end{align}
\end{widetext}

\begin{figure}[t!]
        \centering
        \scalebox{0.6}{
        \begin{tikzpicture}{}
        %circles
        \node[very thick] at (4,6.5) {\Large{Alice}};
        \node[draw, inner sep=25pt](boxa) at (4,4) {};
        \filldraw (4,4) circle (3.5pt);
        \node[very thick](x) at (4,5.3) {\Large{$x_1 \oplus x_2$}};
        \node[very thick](a) at (4,2.7) {\Large{$a$}};
        %arestas

        \node[very thick] at (13,6.5) {\Large{Bob}};
        \node[draw, inner sep=25pt](boxb) at (13,4) {};
        \filldraw (13,4) circle (3.5pt);
        \node[very thick](y) at (13,5.3) {\Large{$y_1 \oplus y_2$}};
        \node[very thick](b) at (13,2.7) {\Large{$b$}};
        %arestas

        \node[very thick] at (8.5,-4) {\Large{Charlie}};
        \node[draw, inner sep=25pt](boxb) at (8.5,-2) {};
        \filldraw (8.5,-2) circle (3.5pt);
        \node[very thick](y) at (8.5,-0.7) {\Large{$z$}};
        \node[very thick](b) at (8.5,-3.3) {\Large{$c$}};

        \node[very thick](mx) at (4,2) {\Large{$m_x = x_1 \oplus a$}};

        \node[very thick](my) at (13,2) {\Large{$m_y = y_1 \oplus b$}};
        
        \draw[very thick, ->] (mx) to[out=285,in=160] (7,-4);
        \draw[very thick, ->] (my) to[out=255,in=20] (10,-4);

         \node[very thick](z0) at (10,-5.5) {\Large{$g_j = m_x \oplus m_y \oplus c$}};
        
        \end{tikzpicture}
        }
        \caption{The communication protocol is performed by Alice, Bob, and Charlie that share a non-signaling resource. Alice (Bob) receive initially two bits $\{x_1, x_2\}$ ($\{y_1, y_2\}$) and perform her local measurements as $x = x_1 \oplus x_2$ ($y =  y_1 \oplus y_2$). After obtaining her outputs $a$ ($b$), encodes the message with $m_x = a\oplus x_1$ ($m_y = b \oplus y_1$). Charlie inputs on his side $z=0$ if he wants to compute $f_1$, and $z=1$ if he wants to compute $f_2$. After receiving the messages, Charlie computes his guess by following $g_j = m_x \oplus m_y \oplus c$.}
        \label{prot2-1}
    \end{figure}
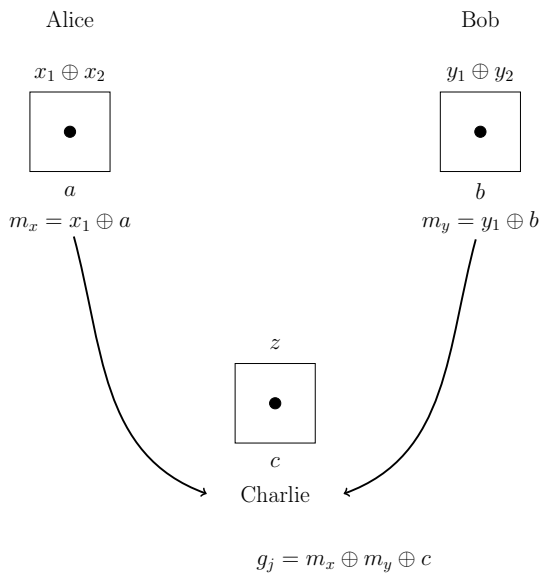

To illustrate, we consider the tripartite scenario, depicted in Fig. \ref{NewScenarioIC}, such that Alice and Bob have just two initial uncorrelated bits and that the communication task of Charlie is to compute two specific functions $f_1 = x^1_1\oplus x^2_1$ and $f_2 = x^1_2\oplus x^2_2$. The communication task is trivialized if the parties share the following tripartite non-signaling (post-quantum) correlation \cite{Pironio_2011},
    \begin{equation}\label{class45}
        p(a,b,c|x,y,z) = \left\{\begin{aligned}
        1/4 & \quad \text{if} \quad a\oplus b \oplus c = xz \oplus yz; \\
        0& \quad\text{else},
        \end{aligned}\right.
    \end{equation}
where $a,b,c,x,y,z \in \{0,1\}$. To achieve it, the parties perform the protocol detailed in Fig. \ref{prot2-1}. In each run, Charlie can always perfectly compute each of the functions, since $g_1 = x^1_1\oplus x^2_1$ and $g_2 = x^1_2\oplus x^2_2$. In other words, similarly to the usual information causality scenario, Charlie has potential access to the four bits of Alice and Bob but receives just two bits communicated by them. Particularizing inequality \eqref{ICmulti} to this case, we obtain
    \begin{multline}\label{triIC}
    \mathcal{I} = I(X_1^1:X_1^2, G_1) + I(X_2^1:X_2^2, G_2)+ \\ + I(X_1^2:X_1^1, G_1) + I(X_2^2:X_2^1, G_2) \le \\ H(M_1, M_2),
    \end{multline}
an inequality that is maximally violated by the NS correlation \eqref{class45} with the protocol described above, since $\mathcal{I} = 4$ while the quantum valid upper bound is $H(M_x,M_y)=2$.
In fact, the multipartite version \eqref{ICmulti} can be violated by the multipartite extension of the post-quantum correlation \eqref{class45}, given by
\begin{widetext}
\begin{equation}\label{multi45}
    p(a_1, a_2, \cdots, a_{N}| x_1, x_2, \cdots, x_{N}) = \left\{\begin{aligned} 1/&2^{N-1}& \quad \text{if} \quad \; &\bigoplus_{k=1}^{N} a_k = \bigoplus_{k=1}^{N-1} x_k x_{N};\\
    &0& \quad \text{else.}& \end{aligned} \right.
\end{equation}
\end{widetext}
Considering $n=2$, $f_j = X_j^1 \oplus X_j^2 \oplus \cdots \oplus X_j^{N-1}$, and the direct extension of the protocol described in Fig. \ref{prot2-1} for the multipartite case, we see that the communication task is trivialized, implying the maximal violation of the multipartite IC inequality \eqref{multi45}.
    
Additionally, it is important to highlight that the multipartite inequality \eqref{ICmulti} does not consist of the parallel application of the criterion \eqref{firstIC} between each sender with the receiver, an approach followed by Ref. \cite{PhysRevA.85.032115}. Indeed, looking at the simplest tripartite case for $n=2$, when the receiver Charlie perfectly computes $g_1 = x^1_1\oplus x^2_1$ and $g_2 = x^1_2\oplus x^2_2$ all informational terms in the left-hand side of Eq.~\eqref{firstIC} vanish, showing that the post-quantum behavior reached with the protocol of Fig.\ref{prot2-1} cannot be detected by this previous approach based on the parallelization of the bipartite IC criterion.
\section{Concatenation procedure}
     \begin{figure*}
        \centering
        \scalebox{0.7}{
        \begin{tikzpicture}{}
        
        \node[very thick](alice) at (-0.2,8.2) {\Large{\textbf{Alice}}};
        
        \node(k1) at (-2.5,6.2) {\Large{$k=2$}};

        \node[very thick](x01) at (0,7.3) {\large{$x_1 \oplus x_2$}};
        \node[fill, draw, inner sep=12pt](boxa01) at (0,6.2) {};
        \node[very thick](a01) at (0,5) {\large{$a_1^2$}};
        \draw[very thick, ->,>=stealth] (x01) -- (boxa01);
        \draw[very thick, ->,>=stealth] (boxa01) -- (a01);

        \node[very thick](x11) at (4,7.3) {\large{$x_3 \oplus x_4$}};
        \node[fill, draw, inner sep=12pt](boxa11) at (4,6.2) {};

        \node[very thick](a11) at (4,5) {\large{$a_2^2$}};
        \draw[very thick, ->,>=stealth] (x11) -- (boxa11);
        \draw[very thick, ->,>=stealth] (boxa11) -- (a11);

        \node(k0) at (-2.5,2.3) {\Large{$k=1$}};

        \node[very thick](x00) at (2,3.9) {\large{$ x_1 \oplus a_1^2 \oplus x_3 \oplus a_2^2$}};
        \node[fill, draw, inner sep=12pt](boxa00) at (2,2.7) {};
        %\node[text width=1cm](j11) at (2,3.5) {$j=0$};
        \node[very thick](a00) at (2,1.5) {\large{$a_1^1$}};
        \draw[very thick, ->,>=stealth] (x00) -- (boxa00);
        \draw[very thick, ->,>=stealth] (boxa00) -- (a00);

        %##########################################
        %######  BOB ##############################

        \node[very thick](y0) at (10,3.9) {\large{$y_1 \oplus b_1^2 \oplus y_3 \oplus b_2^2 $}};
        \node[fill, draw, inner sep=12pt](boxb00) at (10,2.7) {};
        %\node[text width=1cm] at (10,3.5) {$j=0$};
        \node[very thick](b00) at (10,1.5) {\large{$b_1^1$}};
        \draw[very thick, ->,>=stealth] (y0) -- (boxb00);
        \draw[very thick, ->,>=stealth] (boxb00) -- (b00);

        \node[very thick](y1) at (12,7.3) {\large{$y_3 \oplus y_4$}};
        \node[fill, draw, inner sep=12pt](boxb11) at (12,6.2) {};
        %\node[text width=1cm] at (12,7.5) {$j=1$};
        \node[very thick](b11) at (12,5) {\large{$b_2^2$}};
        \draw[very thick, ->,>=stealth] (y1) -- (boxb11);
        \draw[very thick, ->,>=stealth] (boxb11) -- (b11);

        \node[very thick](y00) at (8,7.3) {\large{$y_1 \oplus y_2$}};
        \node[fill, draw, inner sep=12pt](boxb01) at (8,6.2) {};
        %\node[text width=1cm] at (8,7.5) {$j=0$};
        \node[very thick](b01) at (8,5) {$b_1^2$};
        \draw[very thick, ->,>=stealth] (y00) -- (boxb01);
        \draw[very thick, ->,>=stealth] (boxb01) -- (b01);

        \node[very thick](bob) at (7.8,8.2) {\Large{\textbf{Bob}}};

        %##########################################
        %######  CHARLIE ###########################
        
        \node[very thick](z0) at (18,3.9) {\large{$z_1^1$}};
        \node[fill, draw, inner sep=12pt](boxc00) at (18,2.7) {};
        %\node[text width=1cm] at (18,3.5) {$j=0$};
        \node[very thick](c00) at (18,1.5) {\large{$c_1^1$}};
        \draw[very thick, ->,>=stealth] (z0) -- (boxc00);
        \draw[very thick, ->,>=stealth] (boxc00) -- (c00);
        
        \node[black!30!,very thick](z01) at (16,7.3) {\large{$z_1^2$}};
        \node[black!30,fill=black!30, draw, inner sep=12pt](boxc01) at (16,6.2) {};
        %\node[black!30,text width=1cm] at (16,7.5) {$j=0$};
        \node[black!30,very thick](c01) at (16,5) {\large{$c_1^2$}};
        \draw[black!30,very thick, ->,>=stealth] (z01) -- (boxc01);
        \draw[black!30,very thick, ->,>=stealth] (boxc01) -- (c01);
        
        \node[very thick](z1) at (20,7.3) {\large{$z_2^2$}};
        \node[fill, draw, inner sep=12pt](boxc11) at (20,6.2) {};
        %\node[text width=1cm] at (20,7.5) {$j=1$};
        \node[very thick](c11) at (20,5) {\large{$c_2^2$}};
        \draw[very thick, ->,>=stealth] (z1) -- (boxc11);
        \draw[very thick, ->,>=stealth] (boxc11) -- (c11);
        
        \node[very thick](charlie) at (15.8,8.2) {\Large{\textbf{Charlie}}};

        %##########################################
        %######  NIVEIS#############################

        \draw[dashed] (-3.1,4.4) -- (20.5,4.4);

        \end{tikzpicture}}
        \caption{Concatenation performed by Alice, Bob, and Charlie of the protocol in Fig.\ref{prot2-1}. Alice and Bob initially receive $n = 2^K$ bits and Charlie receives a $K$-bitstring $\{z_1,z_2,...,z_{K}\}$, which indicates which pair $x_j \oplus y_j$ he is interested in, $j = \sum_{l=1}^{K} z_l 2^{l-1}$. Thus, Alice and Bob encode their bits in pairs, just following the protocol in Fig.\ref{prot2-1}. Now, instead of sending each respective message, they encode pairs of these in other identical NS-boxes with the same strategy. So, both Alice and Bob perform this procedure until one message remains. Alice and Bob are then allowed to send these one-bit messages to Charlie, which receives the message and to each NS-box performs the decoding protocol just as Fig.\ref{prot2-1}. The idea is that in a given concatenation level $k$, Charlie recovers the sum of Alice and Bob messages previously encoded in the current box, which is associated with a subsequent higher level $k+1$ NS-box. The picture shows a particular case with $n=4$, where $a_i^k$, $b_i^k$, and $c_i^k$ represent the output of the box $i$ in the level $k$ to Alice, Bob and Charlie, respectively. In the level $k=1$ Charlie recovers the messages associated with the box $i=1$ of the level $k=2$ and is able to recover $x_3 \oplus y_3$ or $x_4 \oplus y_4$, depending on $z_2^2$.}
        \label{concatenation}
      \end{figure*}
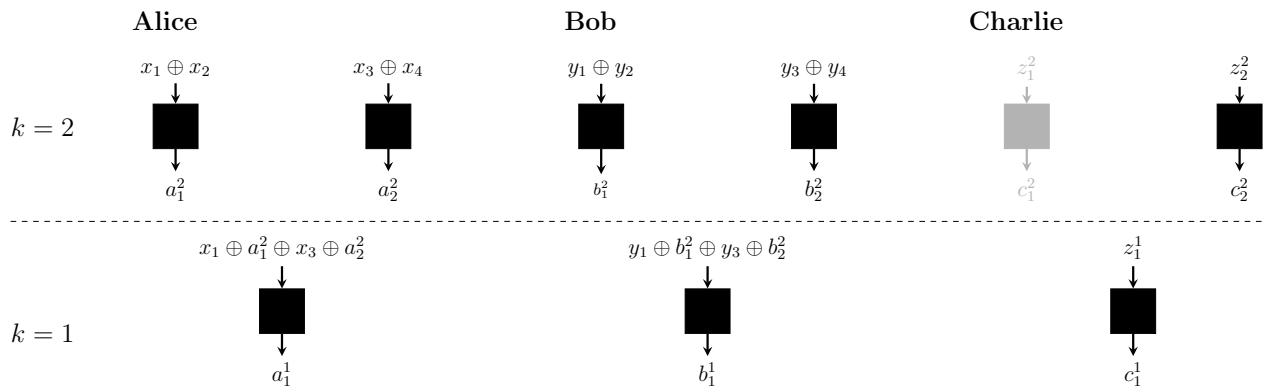
      
As previously discussed, the first proposal for information causality \cite{IC} with the criterion \eqref{firstIC} was able to witness the post-quantum nature of all non-signaling correlations beyond Tsirelson’s bound. For that, however, it was essential to consider a concatenation procedure involving many copies of the correlation under test.  Here we show how such concatenation can be constructed for the tripartite scenario, also generalizing it to arbitrary multipartite scenarios.
        
Similarly to the bipartite scenario, the success probability for the protocol in Fig.\eqref{prot2-1} can be connected to the probability of the resource shared  between the parts, more specifically to the probability $p(a\oplus b \oplus c = xz \oplus yz | x, y, z)$. Clearly, the probabilities of Charlie correctly computing the values of $x_1 \oplus y_1$ and $x_2 \oplus y_2$ are, respectively,
\begin{subequations}\label{pt1pt2}
        \begin{align}
        P_I = \frac{1}{4}[&p(a\oplus b \oplus c = 0|0,0,0) \nonumber\\+ &p(a\oplus b \oplus c = 0|0,1,0) \\+ &p(a\oplus b \oplus c = 0|1,0,0) \nonumber\\ +& p(a\oplus b \oplus c = 0|1,1,0)];\nonumber\end{align}\begin{align}
        P_{II} = \frac{1}{4}[&p(a\oplus b \oplus c = 0|0,0,1)\\ +& p(a\oplus b \oplus c = 1|0,1,1)\nonumber\\ +&p(a\oplus b \oplus c = 1|1,0,1)\nonumber\\ +& p(a\oplus b \oplus c = 0|1,1,1)].\nonumber
        \end{align}
\end{subequations}
In the particular case where the parties share the correlation described by \eqref{class45}, we obtain $P_I = P_{II} = 1$.      
        
In Fig. \ref{concatenation}, we specify the concatenation procedure for the tripartite communication protocol of Fig. \ref{prot2-1}. In this case, Alice and Bob initially receive the respective bit-strings $\mathbf{x}$ and $\mathbf{y}$ of length $n = 2^K$ and share $2^K - 1$ identical copies of binary-input/binary-output non-signaling boxes with Charlie. The success probability that Charlie produces a guess $g_j$ correctly is given by (see appendix \ref{multicommunicationtask})
        \begin{equation}\label{ptsuccess}
            p(g_j = x_j \oplus y_j) =  \frac{1}{2}(1+E_I^{K-r}E_{II}^{r}),
        \end{equation}
where $r$ denotes the number of times that Charlie measures $z=1$ in the $K$ levels of the concatenation code displayed in Fig.\ref{concatenation} and $E_i = 2P_i -1$ (see Eq.\eqref{pt1pt2}). By considering this success probability, we show in Appendix \ref{multiplebits} that information causality is always violated when $E_I^2 + E_{II}^2 > 1$. In other words, when combined with a concatenation procedure and multiple copies of the behaviour under test, the tripartite information causality inequality \eqref{triIC} leads to a generalization of the bipartite inequality \eqref{Uffink}, given by
\begin{equation}
\label{triMultiple} E_I^2 + E_{II}^2 \le 1.
\end{equation}
Similarly to \eqref{triIC}, the multiple copies criterion \eqref{triMultiple} is maximally violated by the behaviour \eqref{class45} since, for this case, $E_I = E_{II} = 1$. Moreover, for isotropic correlations described by a visibility parameter $E$ and such that $E_I = E_{II}= E$, the tripartite multiple copies inequality is violated when $E > 1/\sqrt{2}$, which is exactly the same bound obtained by \cite{IC} for the bipartite scenario. However, for the tripartite scenario, the Navascu\'{e}s-Pironio-Acin (NPA) hierarchy \cite{NPA} implies that for any $E \geq 1/2$ the corresponding correlation will have a post-quantum nature. That is, the tripartite information causality, at least with the specific concatenation considered here, is unable to recover the Tsirelson'sbound.

As previously mentioned, the bipartite version of \eqref{triMultiple} is equivalent to the quadratic constraint obtained by Uffink \cite{UffinkP}. However, for more than two parts, such equivalence no longer holds. For the tripartite scenario, the Uffink inequality reads as
\begin{multline}\label{UffinkTri}
    (C_{001}+C_{010}+C_{100}-C_{111})^2 +  \\
    + (C_{110}+C_{101}+C_{011}-C_{000})^2 \le 16,
\end{multline}
where $C_{xyz} = \sum_{a,b,c} (-1)^{a+b+c}p(a,b,c|x,y,z)$. Indeed, there is no way to alternate between the inequalities \eqref{triMultiple} and \eqref{UffinkTri} by changing labels. Even more importantly, as we will show in the next section, there are post-quantum correlations violating the multiple copies inequality \eqref{triMultiple} that do not violate the tripartite Uffink inequality \eqref{UffinkTri} (and all the inequalities that are obtained from it by relabelling of parties, measurements, and outcomes).
\section{No concatenation version}
The motivation for the generalization of the IC inequality given by \eqref{firstnoisy} comes from the fact that the upper bound in the information gain of the receiver in \eqref{firstIC} should be understood as the single use of a noiseless classical channel of capacity $|M|$. Interestingly, our results in the multipartite scenario can also take into account this insight. Indeed, for the multipartite scenario described in Section \ref{newIC} we may consider that each of the $N-1$ senders performs a single use of a classical noisy channel of capacity $C_k = I(M_k : M_{k}')$, where $M_k$ is the message encoded by each sender $k$ and $M_k '$ is the respective message reaching the decoder after the message passes through the noisy channel. In this case, the criterion defined in \eqref{ICmulti} is easily rewritten in terms of the channel capacity by considering data processing inequalities and the fact that $M_k$ completely determines $M_{k}'$ (that is, $M_{k}'$ is conditionally independent of any random variable $V$ of the causal structure given $M_k$, $I(M_{k}' : V | M_{k}) = 0$). As proven in Appendix \ref{noisyinequality}, it follows that
\begin{multline}\label{ICmultiNoisy}
   \sum_k^{N-1} \sum_i^n I(X_i^k : X_i^1, \dots,X_i^{k-1}, X_i^{k+1}, \dots, X_i^{N-1}, G_i) \\ \le  
    \sum_k^{N-1} C_k + \sum_i^n I(X_{i+1}^k, \dots, X_n^k :  X_i^k).
\end{multline}

Particularizing for the tripartite scenario and, for simplicity, assuming completely uncorrelated initial bits, such a result reads as
\begin{multline}\label{triICmultiplenoisy}
    \mathcal{I} \equiv \sum_{i=1}^{n} [ I(X_i :Y_i , G_i) +  I(Y_i :X_i , G_i ) ] \\ \le I(M_x:M_x ') + I(M_y : M_y ' ),
    \end{multline}
where the two senders have initially $\mathbf{x}^1 = (X_1^k, X_2^k, \cdots, X_n^k)$ and $\mathbf{x}^2 = (Y_1^k, Y_2^k, \cdots, Y_n^k)$, and $M_x '$ and $M_y '$ are the messages reaching the receiver after $M_x$ and $M_y$ pass through the noisy channel, respectively. To illustrate, an application of the noisy tripartite IC inequality will be shown in the next section.

\section{Numerical Tests}

\begin{figure}
        \centering
        \includegraphics[width=1\linewidth,height=0.25\textheight]{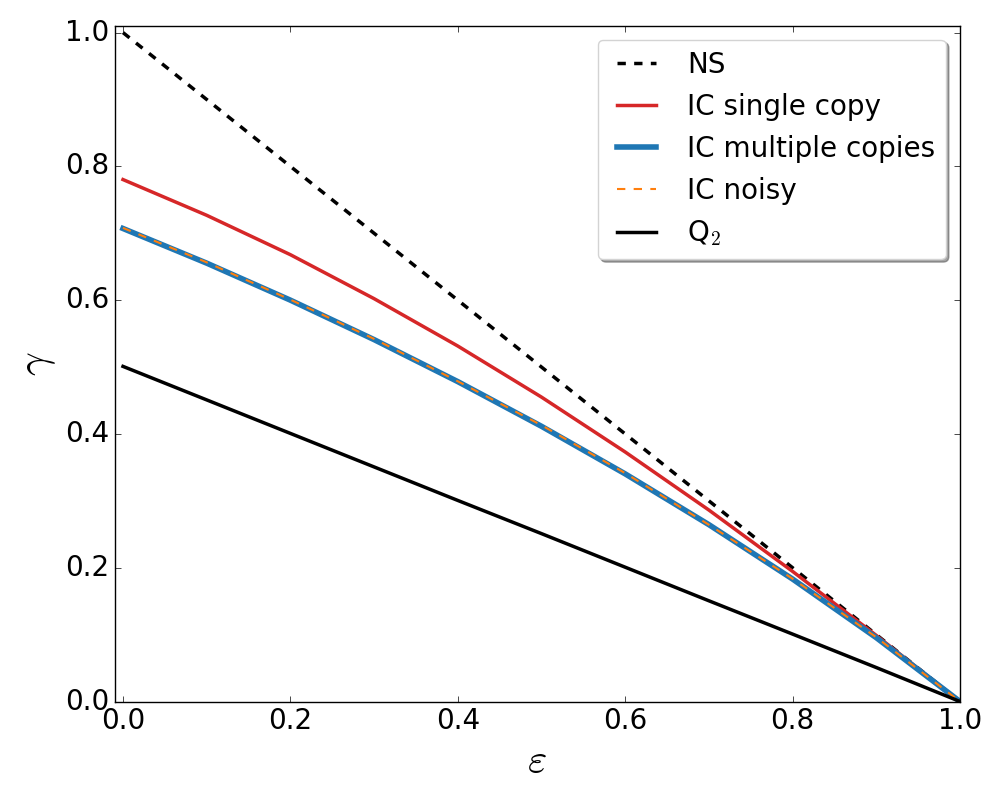}
        \caption{Non-signaling (NS) polytope slice given by Eq.\eqref{eqslice}. Every dot above these curves violates the respective criterion represented. The black dashed and solid lines describe the NS and quantum edges, respectively (the last was computed with the level $2$ of NPA hierarchy \cite{NPA}). The single and multiple copies limits defined by the criteria \eqref{triIC} and \eqref{triMultiple} are respectively depicted with the red and blue solid lines. Finally, the edge defined by the noisy channel criterion \eqref{triICmultiplenoisy} is described by the orange dashed line.}
        \label{slice}
        \end{figure}

More importantly, to understand the strength of the criteria derived, we considered the following slice of the non-signaling set,
        \ba{\label{eqslice}p(a,b,c|x,y,z) = \gamma p_{45} + \epsilon p_D + (1 - \gamma - \epsilon) p_{W},}
        
        \noindent where $\gamma, \epsilon \in [0,1]$, $p_{45}(a,b,c|x,y,z)$ is defined in \eqref{class45}, $p_D(a,b,c|x,y,z) = \delta_{a,0}, \delta_{b,0} \delta_{c,0}$ and $p_W (a,b,c|x,y,z) = 1/8$. Thus, we obtained Fig. \ref{slice}, which highlights that \eqref{triMultiple} excludes even more supra-quantum correlation than \eqref{triIC}.  In addition, despite the distance evidenced between the quantum set and IC, we enforced that the bound \eqref{triMultiple} follows from the particular communication protocol depicted in Fig.\ref{concatenation}. Therefore, it does not exclude the existence of better protocols, able to single out the quantum set for this slice of the non-signaling set or rule out post-quantum extremal correlations.
        
        In Fig.\ref{slice}, also we presented the edge implied by \eqref{triICmultiplenoisy} for the same slice in \eqref{eqslice}, where we considered that all communication is made through a binary symmetric channel that flips the bit with probability $\epsilon$. In this case, to obtain the curve, we followed the results from \cite{ICnoisy} and considered $\epsilon \to 1/2$. From this result, it is clear that our stronger criterion in \eqref{triMultiple} and this new noisy channel approach are in complete agreement, even considering the simplest noisy channel. The codes related to the Fig. \ref{slice} are available in \cite{Slices}.
        
        For the tripartite scenario with binary-input/binary-output there exist, 53856 non-signaling extremal correlations that are divided into 46 different equivalence classes, among which 45 are supra-quantum ones \cite{Pironio_2011}. Thus, we also checked the ability of \eqref{triMultiple} to exclude all the supra-quantum extremals of the non-signaling set. In Table \ref{violate}, we highlight those classes for which we could find a violation of \eqref{triMultiple}. Furthermore, Table \ref{violate} contains the same analysis for the tripartite Uffink inequality \eqref{UffinkTri}. From these results, it is clear the no equivalence between the multiple copies IC inequality \eqref{triMultiple} and the Uffink result \eqref{UffinkTri}, since there exist extremal non-signaling correlations which respect one constraint while violating the other. The codes related to the results from Table \ref{violate} are available in \cite{ExtremalViolation}.
        \begin{table}
        \caption{\label{violate}Classes of non-signaling extremal correlations defined in \cite{Pironio_2011} that violate \eqref{triMultiple} or \eqref{UffinkTri}}
		\begin{ruledtabular}
		\begin{tabular}{lcr}
		\textrm{Inequality}&
		\textrm{Extremal boxes}&\\
		\colrule
		\eqref{triMultiple} & 35, 37, 38, 40, 41, 42, 43, 44, 45 &\\
		\eqref{UffinkTri} & 21, 22, 30, 34, 36, 39, 41, 44, 46 &\\
		\end{tabular}
		\end{ruledtabular}
		\end{table}

\section{Conclusion}

We proposed a new multipartite communication task in which the previous IC formulation does not detect non-local advantage. Thus, by employing the quantum causal structure formalism, we proposed a new criterion to describe IC in such a new context and proved its truthfulness for the whole set of quantum correlations, for any number of parts. Furthermore, we proved that our model allows the concatenation approach from \cite{IC}, which enabled us to derive even stronger constraints for the multipartite non-signaling correlations set. In that case, our multipartite inequality proved to be strictly stronger than the multipartite Uffink's inequality from \cite{UffinkP}, which contrasts with the previous bipartite result from \cite{IC}. In addition, our findings are in complete agreement with the recent noisy channel approach from \cite{ICnoisy}, which allows many other analyses for such a multipartite context.

We emphasize that our results are limited by one specific protocol, which is optimal to Eq.\eqref{multi45}, however, it does not ensure that it is optimal for all non-signaling correlations. Thus, searching for better protocols for different correlations may yield stronger results. Furthermore, the analysis of non-dichotomy scenarios, or even cases where the sender's initial bits are correlated, may also produce interesting results, as previously analyzed in \cite{MLviolatesIC}. Moreover, our findings open a new class of non-sequential multipartite RACs, where multiple parts send messages to each other with the task to compute a boolean function of the senders' initial bits. The figure of merit, in this case, is to compute the success probability concerning the receiver to compute correctly such a function. Thus, investigating these new thresholds for such a probability of success may have important implications for quantum information processing.

\begin{acknowledgments}
    We thank Marcelo Terra Cunha and Pedro Lauand for fruitful discussions and suggestions. We especially acknowledge Pedro Lauand for providing the extremal non-signaling behaviors leading to the results in Table \ref{violate}.
    This work was also supported by the Brazilian National Council for Scientific and Technological Development (CNPq, grant No. 307295/2020-6), the National Institute for Science and Technology on Quantum Information (INCT-IQ) (Grant No. 465469/2014-0),  the Serrapilheira Institute (Grant No. Serra-1708-15763) and the São Paulo Research Foundation FAPESP (Grant No. 2018/07258-7 and 2019/00451-9).
\end{acknowledgments}

\bibliography{ref.bib}

\begin{thebibliography}{26}%
\makeatletter
\providecommand \@ifxundefined [1]{%
 \@ifx{#1\undefined}
}%
\providecommand \@ifnum [1]{%
 \ifnum #1\expandafter \@firstoftwo
 \else \expandafter \@secondoftwo
 \fi
}%
\providecommand \@ifx [1]{%
 \ifx #1\expandafter \@firstoftwo
 \else \expandafter \@secondoftwo
 \fi
}%
\providecommand \natexlab [1]{#1}%
\providecommand \enquote  [1]{``#1''}%
\providecommand \bibnamefont  [1]{#1}%
\providecommand \bibfnamefont [1]{#1}%
\providecommand \citenamefont [1]{#1}%
\providecommand \href@noop [0]{\@secondoftwo}%
\providecommand \href [0]{\begingroup \@sanitize@url \@href}%
\providecommand \@href[1]{\@@startlink{#1}\@@href}%
\providecommand \@@href[1]{\endgroup#1\@@endlink}%
\providecommand \@sanitize@url [0]{\catcode `\\12\catcode `\$12\catcode
  `\&12\catcode `\#12\catcode `\^12\catcode `\_12\catcode `\%12\relax}%
\providecommand \@@startlink[1]{}%
\providecommand \@@endlink[0]{}%
\providecommand \url  [0]{\begingroup\@sanitize@url \@url }%
\providecommand \@url [1]{\endgroup\@href {#1}{\urlprefix }}%
\providecommand \urlprefix  [0]{URL }%
\providecommand \Eprint [0]{\href }%
\providecommand \doibase [0]{https://doi.org/}%
\providecommand \selectlanguage [0]{\@gobble}%
\providecommand \bibinfo  [0]{\@secondoftwo}%
\providecommand \bibfield  [0]{\@secondoftwo}%
\providecommand \translation [1]{[#1]}%
\providecommand \BibitemOpen [0]{}%
\providecommand \bibitemStop [0]{}%
\providecommand \bibitemNoStop [0]{.\EOS\space}%
\providecommand \EOS [0]{\spacefactor3000\relax}%
\providecommand \BibitemShut  [1]{\csname bibitem#1\endcsname}%
\let\auto@bib@innerbib\@empty
%</preamble>
\bibitem [{\citenamefont {Bell}(1964)}]{Bell1964}%
  \BibitemOpen
  \bibfield  {author} {\bibinfo {author} {\bibfnamefont {J.~S.}\ \bibnamefont
  {Bell}},\ }\bibfield  {title} {\bibinfo {title} {On the einstein podolsky
  rosen paradox},\ }\href@noop {} {\bibfield  {journal} {\bibinfo  {journal}
  {Physics Physique Fizika}\ ,\ \bibinfo {pages} {195}} (\bibinfo {year}
  {1964})}\BibitemShut {NoStop}%
\bibitem [{\citenamefont {Brunner}\ \emph {et~al.}(2014)\citenamefont
  {Brunner}, \citenamefont {Cavalcanti}, \citenamefont {Pironio}, \citenamefont
  {Scarani},\ and\ \citenamefont {Wehner}}]{Brunner_etal2014}%
  \BibitemOpen
  \bibfield  {author} {\bibinfo {author} {\bibfnamefont {N.}~\bibnamefont
  {Brunner}}, \bibinfo {author} {\bibfnamefont {D.}~\bibnamefont {Cavalcanti}},
  \bibinfo {author} {\bibfnamefont {S.}~\bibnamefont {Pironio}}, \bibinfo
  {author} {\bibfnamefont {V.}~\bibnamefont {Scarani}},\ and\ \bibinfo {author}
  {\bibfnamefont {S.}~\bibnamefont {Wehner}},\ }\bibfield  {title} {\bibinfo
  {title} {Bell nonlocality},\ }\href@noop {} {\bibfield  {journal} {\bibinfo
  {journal} {Reviews of Modern Physics}\ ,\ \bibinfo {pages} {419}} (\bibinfo
  {year} {2014})}\BibitemShut {NoStop}%
\bibitem [{\citenamefont {Cirel'son}(1980{\natexlab{a}})}]{Tsirelson_lim}%
  \BibitemOpen
  \bibfield  {author} {\bibinfo {author} {\bibfnamefont {B.~S.}\ \bibnamefont
  {Cirel'son}},\ }\bibfield  {title} {\bibinfo {title} {Quantum generalizations
  of bell's inequality},\ }\href {https://doi.org/10.1007/BF00417500}
  {\bibfield  {journal} {\bibinfo  {journal} {Letters in Mathematical Physics}\
  }\textbf {\bibinfo {volume} {4}},\ \bibinfo {pages} {93} (\bibinfo {year}
  {1980}{\natexlab{a}})}\BibitemShut {NoStop}%
\bibitem [{\citenamefont {Popescu}\ and\ \citenamefont
  {Rohrlich}(1994)}]{PRbox}%
  \BibitemOpen
  \bibfield  {author} {\bibinfo {author} {\bibfnamefont {S.}~\bibnamefont
  {Popescu}}\ and\ \bibinfo {author} {\bibfnamefont {D.}~\bibnamefont
  {Rohrlich}},\ }\bibfield  {title} {\bibinfo {title} {Quantum nonlocality as
  an axiom.},\ }\href {https://doi.org/10.1007/BF02058098} {\bibfield
  {journal} {\bibinfo  {journal} {Foundations of Physics}\ }\textbf {\bibinfo
  {volume} {24}},\ \bibinfo {pages} {379} (\bibinfo {year} {1994})}\BibitemShut
  {NoStop}%
\bibitem [{\citenamefont {Brassard}\ \emph {et~al.}(2006)\citenamefont
  {Brassard}, \citenamefont {Buhrman}, \citenamefont {Linden}, \citenamefont
  {M\'ethot}, \citenamefont {Tapp},\ and\ \citenamefont
  {Unger}}]{NontrivialCC}%
  \BibitemOpen
  \bibfield  {author} {\bibinfo {author} {\bibfnamefont {G.}~\bibnamefont
  {Brassard}}, \bibinfo {author} {\bibfnamefont {H.}~\bibnamefont {Buhrman}},
  \bibinfo {author} {\bibfnamefont {N.}~\bibnamefont {Linden}}, \bibinfo
  {author} {\bibfnamefont {A.~A.}\ \bibnamefont {M\'ethot}}, \bibinfo {author}
  {\bibfnamefont {A.}~\bibnamefont {Tapp}},\ and\ \bibinfo {author}
  {\bibfnamefont {F.}~\bibnamefont {Unger}},\ }\bibfield  {title} {\bibinfo
  {title} {Limit on nonlocality in any world in which communication complexity
  is not trivial},\ }\href {https://doi.org/10.1103/PhysRevLett.96.250401}
  {\bibfield  {journal} {\bibinfo  {journal} {Phys. Rev. Lett.}\ }\textbf
  {\bibinfo {volume} {96}},\ \bibinfo {pages} {250401} (\bibinfo {year}
  {2006})}\BibitemShut {NoStop}%
\bibitem [{\citenamefont {Navascu{\'e}s}\ and\ \citenamefont
  {Wunderlich}(2009)}]{ML}%
  \BibitemOpen
  \bibfield  {author} {\bibinfo {author} {\bibfnamefont {M.}~\bibnamefont
  {Navascu{\'e}s}}\ and\ \bibinfo {author} {\bibfnamefont {H.}~\bibnamefont
  {Wunderlich}},\ }\bibfield  {title} {\bibinfo {title} {A glance beyond the
  quantum model},\ }\href {https://doi.org/10.1098/rspa.2009.0453} {\bibfield
  {journal} {\bibinfo  {journal} {Proceedings of the Royal Society A:
  Mathematical, Physical and Engineering Sciences}\ }\textbf {\bibinfo {volume}
  {466}},\ \bibinfo {pages} {881} (\bibinfo {year} {2009})}\BibitemShut
  {NoStop}%
\bibitem [{\citenamefont {Fritz}\ \emph {et~al.}(2013)\citenamefont {Fritz},
  \citenamefont {Sainz}, \citenamefont {Augusiak}, \citenamefont {Brask},
  \citenamefont {Chaves}, \citenamefont {Leverrier},\ and\ \citenamefont
  {Ac{\'\i}n}}]{OL}%
  \BibitemOpen
  \bibfield  {author} {\bibinfo {author} {\bibfnamefont {T.}~\bibnamefont
  {Fritz}}, \bibinfo {author} {\bibfnamefont {A.}~\bibnamefont {Sainz}},
  \bibinfo {author} {\bibfnamefont {R.}~\bibnamefont {Augusiak}}, \bibinfo
  {author} {\bibfnamefont {J.~B.}\ \bibnamefont {Brask}}, \bibinfo {author}
  {\bibfnamefont {R.}~\bibnamefont {Chaves}}, \bibinfo {author} {\bibfnamefont
  {A.}~\bibnamefont {Leverrier}},\ and\ \bibinfo {author} {\bibfnamefont
  {A.}~\bibnamefont {Ac{\'\i}n}},\ }\bibfield  {title} {\bibinfo {title} {Local
  orthogonality as a multipartite principle for quantum correlations},\
  }\bibfield  {journal} {\bibinfo  {journal} {Nature Communications}\ }\textbf
  {\bibinfo {volume} {4}},\ \href {https://doi.org/10.1038/ncomms3263}
  {10.1038/ncomms3263} (\bibinfo {year} {2013})\BibitemShut {NoStop}%
\bibitem [{\citenamefont {Navascu{\'e}s}\ \emph {et~al.}(2015)\citenamefont
  {Navascu{\'e}s}, \citenamefont {Guryanova}, \citenamefont {Hoban},\ and\
  \citenamefont {Ac{\'\i}n}}]{navascues2015almost}%
  \BibitemOpen
  \bibfield  {author} {\bibinfo {author} {\bibfnamefont {M.}~\bibnamefont
  {Navascu{\'e}s}}, \bibinfo {author} {\bibfnamefont {Y.}~\bibnamefont
  {Guryanova}}, \bibinfo {author} {\bibfnamefont {M.~J.}\ \bibnamefont
  {Hoban}},\ and\ \bibinfo {author} {\bibfnamefont {A.}~\bibnamefont
  {Ac{\'\i}n}},\ }\bibfield  {title} {\bibinfo {title} {Almost quantum
  correlations},\ }\href {https://doi.org/10.1038/ncomms7288} {\bibfield
  {journal} {\bibinfo  {journal} {Nature communications}\ }\textbf {\bibinfo
  {volume} {6}},\ \bibinfo {pages} {1} (\bibinfo {year} {2015})}\BibitemShut
  {NoStop}%
\bibitem [{\citenamefont {Paw{\l}owski}\ \emph {et~al.}(2009)\citenamefont
  {Paw{\l}owski}, \citenamefont {Paterek}, \citenamefont {Kaszlikowski},
  \citenamefont {Scarani}, \citenamefont {Winter},\ and\ \citenamefont
  {{\.Z}ukowski}}]{IC}%
  \BibitemOpen
  \bibfield  {author} {\bibinfo {author} {\bibfnamefont {M.}~\bibnamefont
  {Paw{\l}owski}}, \bibinfo {author} {\bibfnamefont {T.}~\bibnamefont
  {Paterek}}, \bibinfo {author} {\bibfnamefont {D.}~\bibnamefont
  {Kaszlikowski}}, \bibinfo {author} {\bibfnamefont {V.}~\bibnamefont
  {Scarani}}, \bibinfo {author} {\bibfnamefont {A.}~\bibnamefont {Winter}},\
  and\ \bibinfo {author} {\bibfnamefont {M.}~\bibnamefont {{\.Z}ukowski}},\
  }\bibfield  {title} {\bibinfo {title} {Information causality as a physical
  principle},\ }\href {https://doi.org/10.1038/nature08400} {\bibfield
  {journal} {\bibinfo  {journal} {Nature}\ }\textbf {\bibinfo {volume} {461}},\
  \bibinfo {pages} {1101} (\bibinfo {year} {2009})}\BibitemShut {NoStop}%
\bibitem [{\citenamefont {Allcock}\ \emph {et~al.}(2009)\citenamefont
  {Allcock}, \citenamefont {Brunner}, \citenamefont {Pawlowski},\ and\
  \citenamefont {Scarani}}]{Allcock_2009}%
  \BibitemOpen
  \bibfield  {author} {\bibinfo {author} {\bibfnamefont {J.}~\bibnamefont
  {Allcock}}, \bibinfo {author} {\bibfnamefont {N.}~\bibnamefont {Brunner}},
  \bibinfo {author} {\bibfnamefont {M.}~\bibnamefont {Pawlowski}},\ and\
  \bibinfo {author} {\bibfnamefont {V.}~\bibnamefont {Scarani}},\ }\bibfield
  {title} {\bibinfo {title} {Recovering part of the boundary between quantum
  and nonquantum correlations from information causality},\ }\bibfield
  {journal} {\bibinfo  {journal} {Physical Review A}\ }\textbf {\bibinfo
  {volume} {80}},\ \href {https://doi.org/10.1103/physreva.80.040103}
  {10.1103/physreva.80.040103} (\bibinfo {year} {2009})\BibitemShut {NoStop}%
\bibitem [{\citenamefont {Chaves}\ \emph {et~al.}(2015)\citenamefont {Chaves},
  \citenamefont {Majenz},\ and\ \citenamefont {Gross}}]{Chaves_2015}%
  \BibitemOpen
  \bibfield  {author} {\bibinfo {author} {\bibfnamefont {R.}~\bibnamefont
  {Chaves}}, \bibinfo {author} {\bibfnamefont {C.}~\bibnamefont {Majenz}},\
  and\ \bibinfo {author} {\bibfnamefont {D.}~\bibnamefont {Gross}},\ }\bibfield
   {title} {\bibinfo {title} {Information--theoretic implications of quantum
  causal structures},\ }\bibfield  {journal} {\bibinfo  {journal} {Nature
  Communications}\ }\textbf {\bibinfo {volume} {6}},\ \href
  {https://doi.org/10.1038/ncomms6766} {10.1038/ncomms6766} (\bibinfo {year}
  {2015})\BibitemShut {NoStop}%
\bibitem [{\citenamefont {Yu}\ and\ \citenamefont
  {Scarani}(2022)}]{yu2022information}%
  \BibitemOpen
  \bibfield  {author} {\bibinfo {author} {\bibfnamefont {B.}~\bibnamefont
  {Yu}}\ and\ \bibinfo {author} {\bibfnamefont {V.}~\bibnamefont {Scarani}},\
  }\href@noop {} {\bibinfo {title} {Information causality beyond the random
  access code model}} (\bibinfo {year} {2022}),\ \Eprint
  {https://arxiv.org/abs/2201.08986} {arXiv:2201.08986 [quant-ph]} \BibitemShut
  {NoStop}%
\bibitem [{\citenamefont {Gallego}\ \emph {et~al.}(2011)\citenamefont
  {Gallego}, \citenamefont {W\"urflinger}, \citenamefont {Ac\'{\i}n},\ and\
  \citenamefont {Navascu\'es}}]{PhysRevLett.107.210403}%
  \BibitemOpen
  \bibfield  {author} {\bibinfo {author} {\bibfnamefont {R.}~\bibnamefont
  {Gallego}}, \bibinfo {author} {\bibfnamefont {L.~E.}\ \bibnamefont
  {W\"urflinger}}, \bibinfo {author} {\bibfnamefont {A.}~\bibnamefont
  {Ac\'{\i}n}},\ and\ \bibinfo {author} {\bibfnamefont {M.}~\bibnamefont
  {Navascu\'es}},\ }\bibfield  {title} {\bibinfo {title} {Quantum correlations
  require multipartite information principles},\ }\href
  {https://doi.org/10.1103/PhysRevLett.107.210403} {\bibfield  {journal}
  {\bibinfo  {journal} {Phys. Rev. Lett.}\ }\textbf {\bibinfo {volume} {107}},\
  \bibinfo {pages} {210403} (\bibinfo {year} {2011})}\BibitemShut {NoStop}%
\bibitem [{\citenamefont {Yang}\ \emph {et~al.}(2012)\citenamefont {Yang},
  \citenamefont {Cavalcanti}, \citenamefont {Almeida}, \citenamefont {Teo},\
  and\ \citenamefont {Scarani}}]{Yang_2012}%
  \BibitemOpen
  \bibfield  {author} {\bibinfo {author} {\bibfnamefont {T.~H.}\ \bibnamefont
  {Yang}}, \bibinfo {author} {\bibfnamefont {D.}~\bibnamefont {Cavalcanti}},
  \bibinfo {author} {\bibfnamefont {M.~L.}\ \bibnamefont {Almeida}}, \bibinfo
  {author} {\bibfnamefont {C.}~\bibnamefont {Teo}},\ and\ \bibinfo {author}
  {\bibfnamefont {V.}~\bibnamefont {Scarani}},\ }\bibfield  {title} {\bibinfo
  {title} {Information-causality and extremal tripartite correlations},\ }\href
  {https://doi.org/10.1088/1367-2630/14/1/013061} {\bibfield  {journal}
  {\bibinfo  {journal} {New Journal of Physics}\ }\textbf {\bibinfo {volume}
  {14}},\ \bibinfo {pages} {013061} (\bibinfo {year} {2012})}\BibitemShut
  {NoStop}%
\bibitem [{\citenamefont {Miklin}\ and\ \citenamefont
  {Paw\l{}owski}(2021)}]{ICnoisy}%
  \BibitemOpen
  \bibfield  {author} {\bibinfo {author} {\bibfnamefont {N.}~\bibnamefont
  {Miklin}}\ and\ \bibinfo {author} {\bibfnamefont {M.}~\bibnamefont
  {Paw\l{}owski}},\ }\bibfield  {title} {\bibinfo {title} {Information
  causality without concatenation},\ }\href
  {https://doi.org/10.1103/PhysRevLett.126.220403} {\bibfield  {journal}
  {\bibinfo  {journal} {Phys. Rev. Lett.}\ }\textbf {\bibinfo {volume} {126}},\
  \bibinfo {pages} {220403} (\bibinfo {year} {2021})}\BibitemShut {NoStop}%
\bibitem [{\citenamefont {Uffink}(2002)}]{UffinkP}%
  \BibitemOpen
  \bibfield  {author} {\bibinfo {author} {\bibfnamefont {J.}~\bibnamefont
  {Uffink}},\ }\bibfield  {title} {\bibinfo {title} {Quadratic bell
  inequalities as tests for multipartite entanglement},\ }\href
  {https://doi.org/10.1103/PhysRevLett.88.230406} {\bibfield  {journal}
  {\bibinfo  {journal} {Phys. Rev. Lett.}\ }\textbf {\bibinfo {volume} {88}},\
  \bibinfo {pages} {230406} (\bibinfo {year} {2002})}\BibitemShut {NoStop}%
\bibitem [{\citenamefont {Clauser}\ \emph {et~al.}(1969)\citenamefont
  {Clauser}, \citenamefont {Horne}, \citenamefont {Shimony},\ and\
  \citenamefont {Holt}}]{CHSH}%
  \BibitemOpen
  \bibfield  {author} {\bibinfo {author} {\bibfnamefont {J.~F.}\ \bibnamefont
  {Clauser}}, \bibinfo {author} {\bibfnamefont {M.~A.}\ \bibnamefont {Horne}},
  \bibinfo {author} {\bibfnamefont {A.}~\bibnamefont {Shimony}},\ and\ \bibinfo
  {author} {\bibfnamefont {R.~A.}\ \bibnamefont {Holt}},\ }\bibfield  {title}
  {\bibinfo {title} {Proposed experiment to test local hidden-variable
  theories},\ }\href {https://doi.org/10.1103/PhysRevLett.23.880} {\bibfield
  {journal} {\bibinfo  {journal} {Phys. Rev. Lett.}\ }\textbf {\bibinfo
  {volume} {23}},\ \bibinfo {pages} {880} (\bibinfo {year} {1969})}\BibitemShut
  {NoStop}%
\bibitem [{\citenamefont {Cirel'son}(1980{\natexlab{b}})}]{cirel1980quantum}%
  \BibitemOpen
  \bibfield  {author} {\bibinfo {author} {\bibfnamefont {B.~S.}\ \bibnamefont
  {Cirel'son}},\ }\bibfield  {title} {\bibinfo {title} {Quantum generalizations
  of bell's inequality},\ }\href {https://doi.org/10.1007/BF00417500}
  {\bibfield  {journal} {\bibinfo  {journal} {Letters in Mathematical Physics}\
  }\textbf {\bibinfo {volume} {4}},\ \bibinfo {pages} {93} (\bibinfo {year}
  {1980}{\natexlab{b}})}\BibitemShut {NoStop}%
\bibitem [{\citenamefont {Brito}\ \emph {et~al.}(2019)\citenamefont {Brito},
  \citenamefont {Moreno}, \citenamefont {Rai},\ and\ \citenamefont
  {Chaves}}]{brito2019nonlocality}%
  \BibitemOpen
  \bibfield  {author} {\bibinfo {author} {\bibfnamefont {S.~G.~A.}\
  \bibnamefont {Brito}}, \bibinfo {author} {\bibfnamefont {M.~G.~M.}\
  \bibnamefont {Moreno}}, \bibinfo {author} {\bibfnamefont {A.}~\bibnamefont
  {Rai}},\ and\ \bibinfo {author} {\bibfnamefont {R.}~\bibnamefont {Chaves}},\
  }\bibfield  {title} {\bibinfo {title} {Nonlocality distillation and quantum
  voids},\ }\href {https://doi.org/10.1103/PhysRevA.100.012102} {\bibfield
  {journal} {\bibinfo  {journal} {Phys. Rev. A}\ }\textbf {\bibinfo {volume}
  {100}},\ \bibinfo {pages} {012102} (\bibinfo {year} {2019})}\BibitemShut
  {NoStop}%
\bibitem [{\citenamefont {Rai}\ \emph {et~al.}(2019)\citenamefont {Rai},
  \citenamefont {Duarte}, \citenamefont {Brito},\ and\ \citenamefont
  {Chaves}}]{rai2019geometry}%
  \BibitemOpen
  \bibfield  {author} {\bibinfo {author} {\bibfnamefont {A.}~\bibnamefont
  {Rai}}, \bibinfo {author} {\bibfnamefont {C.}~\bibnamefont {Duarte}},
  \bibinfo {author} {\bibfnamefont {S.}~\bibnamefont {Brito}},\ and\ \bibinfo
  {author} {\bibfnamefont {R.}~\bibnamefont {Chaves}},\ }\bibfield  {title}
  {\bibinfo {title} {Geometry of the quantum set on no-signaling faces},\
  }\href {https://doi.org/10.1103/PhysRevA.99.032106} {\bibfield  {journal}
  {\bibinfo  {journal} {Phys. Rev. A}\ }\textbf {\bibinfo {volume} {99}},\
  \bibinfo {pages} {032106} (\bibinfo {year} {2019})}\BibitemShut {NoStop}%
\bibitem [{\citenamefont {Pironio}\ \emph {et~al.}(2011)\citenamefont
  {Pironio}, \citenamefont {Bancal},\ and\ \citenamefont
  {Scarani}}]{Pironio_2011}%
  \BibitemOpen
  \bibfield  {author} {\bibinfo {author} {\bibfnamefont {S.}~\bibnamefont
  {Pironio}}, \bibinfo {author} {\bibfnamefont {J.-D.}\ \bibnamefont
  {Bancal}},\ and\ \bibinfo {author} {\bibfnamefont {V.}~\bibnamefont
  {Scarani}},\ }\bibfield  {title} {\bibinfo {title} {Extremal correlations of
  the tripartite no-signaling polytope},\ }\href
  {https://doi.org/10.1088/1751-8113/44/6/065303} {\bibfield  {journal}
  {\bibinfo  {journal} {Journal of Physics A: Mathematical and Theoretical}\
  }\textbf {\bibinfo {volume} {44}},\ \bibinfo {pages} {065303} (\bibinfo
  {year} {2011})}\BibitemShut {NoStop}%
\bibitem [{\citenamefont {Hsu}(2012)}]{PhysRevA.85.032115}%
  \BibitemOpen
  \bibfield  {author} {\bibinfo {author} {\bibfnamefont {L.-Y.}\ \bibnamefont
  {Hsu}},\ }\bibfield  {title} {\bibinfo {title} {Multipartite information
  causality},\ }\href {https://doi.org/10.1103/PhysRevA.85.032115} {\bibfield
  {journal} {\bibinfo  {journal} {Phys. Rev. A}\ }\textbf {\bibinfo {volume}
  {85}},\ \bibinfo {pages} {032115} (\bibinfo {year} {2012})}\BibitemShut
  {NoStop}%
\bibitem [{\citenamefont {Navascu{\'e}s}\ \emph {et~al.}(2008)\citenamefont
  {Navascu{\'e}s}, \citenamefont {Pironio},\ and\ \citenamefont
  {Ac{\'\i}n}}]{NPA}%
  \BibitemOpen
  \bibfield  {author} {\bibinfo {author} {\bibfnamefont {M.}~\bibnamefont
  {Navascu{\'e}s}}, \bibinfo {author} {\bibfnamefont {S.}~\bibnamefont
  {Pironio}},\ and\ \bibinfo {author} {\bibfnamefont {A.}~\bibnamefont
  {Ac{\'\i}n}},\ }\bibfield  {title} {\bibinfo {title} {A convergent hierarchy
  of semidefinite programs characterizing the set of quantum correlations},\
  }\href {https://doi.org/10.1088/1367-2630/10/7/073013} {\bibfield  {journal}
  {\bibinfo  {journal} {New Journal of Physics}\ }\textbf {\bibinfo {volume}
  {10}},\ \bibinfo {pages} {073013} (\bibinfo {year} {2008})}\BibitemShut
  {NoStop}%
\bibitem [{\citenamefont {Pollyceno}(2022{\natexlab{a}})}]{Slices}%
  \BibitemOpen
  \bibfield  {author} {\bibinfo {author} {\bibfnamefont {L.}~\bibnamefont
  {Pollyceno}},\ }\href {https://github.com/Pollyceno/TripartiteSlices}
  {\bibinfo {title} {Code concerning the figure \ref{slice}}} (\bibinfo {year}
  {2022}{\natexlab{a}})\BibitemShut {NoStop}%
\bibitem [{\citenamefont {Pollyceno}(2022{\natexlab{b}})}]{ExtremalViolation}%
  \BibitemOpen
  \bibfield  {author} {\bibinfo {author} {\bibfnamefont {L.}~\bibnamefont
  {Pollyceno}},\ }\href {https://github.com/Pollyceno/TripartiteViolation}
  {\bibinfo {title} {Code concerning the results from the table \ref{violate}}}
  (\bibinfo {year} {2022}{\natexlab{b}})\BibitemShut {NoStop}%
\bibitem [{\citenamefont {Cavalcanti}\ \emph {et~al.}(2010)\citenamefont
  {Cavalcanti}, \citenamefont {Salles},\ and\ \citenamefont
  {Scarani}}]{MLviolatesIC}%
  \BibitemOpen
  \bibfield  {author} {\bibinfo {author} {\bibfnamefont {D.}~\bibnamefont
  {Cavalcanti}}, \bibinfo {author} {\bibfnamefont {A.}~\bibnamefont {Salles}},\
  and\ \bibinfo {author} {\bibfnamefont {V.}~\bibnamefont {Scarani}},\
  }\bibfield  {title} {\bibinfo {title} {Macroscopically local correlations can
  violate information causality},\ }\bibfield  {journal} {\bibinfo  {journal}
  {Nature Communications}\ }\textbf {\bibinfo {volume} {1}},\ \href
  {https://doi.org/10.1038/ncomms1138} {10.1038/ncomms1138} (\bibinfo {year}
  {2010})\BibitemShut {NoStop}%
\end{thebibliography}%

%\end{multicols}{2}

\newpage

\onecolumngrid

\appendix

\section{Concatenation in a multipartite communication task}\label{multicommunicationtask}

Here we extend the tripartite communication task from section \ref{newIC} to a general multipartite scenario. Thus, consider $N$ parts, among which $N-1$ are senders that initially have their respective bit-strings $\mathbf{x}^k = (X_1^k, X_2^k, \cdots, X_n^k)$, where $k\in \{1, 2, \cdots, N-1\}$. Each sender encodes a classical message $M_k$ of size $m < n$ to the $N^{th}$-part, the receiver. This last one needs rightly compute one of $n$ possible initial bits functions $f_j (X_j^1, X_j^2, \cdots, X_j^{N-1})$, by producing the guess $G_j$, where $j\in \{1, \cdots, n\}$. Just as in the main text, in addition to the classical messages, non-signaling correlations are allowed among all $N$ parts.

Now consider a little more particular case, where $n=2$ and $f_j = X_j^1 \oplus X_j^2 \oplus \cdots \oplus X_j^{N-1}$. Just as in the previously described tripartite scenario, we find such a particular multipartite communication task is trivialized by a generalization of the correlation \eqref{class45} for the $(N, 2, 2)$ Bell scenario, \ie

\begin{equation}
    p(a_1, a_2, \cdots, a_{N}| x_1, x_2, \cdots, x_{N}) = \left\{\begin{aligned} 1/&2^{N-1}& \quad \text{if} \quad \; &\bigoplus_{k=1}^{N} a_k = \bigoplus_{k=1}^{N-1} x_k x_{N};\\
    &0& \quad \text{else.}& \end{aligned} \right.
\end{equation}

\noindent where $a_k$ and $x_k$ respectively denote the output and input of the part $k$. To see this, consider that the $N$ parts perform the strategy depicted in Fig.\ref{prot2-1}. That is, each sender performs the encoding $x_k = X_1^k \oplus X_2^k$ and $M_k = X_1^k \oplus a_k$, and the receiver computes the guess $G_j = \bigoplus_{k=1}^{N-1} M_k \oplus a_N$. In this case, by considering \eqref{multi45} we find

\begin{align}
    G_j &= \bigoplus_{k=1}^{N-1} ( X_1^k \oplus a_k ) \oplus a_N;\nonumber \\
    &= \left(\bigoplus_{k=1}^{N-1} X_1^k \right)\oplus \left(\bigoplus_{k=1}^{N} a_k\right) \nonumber\\
    &= \left(\bigoplus_{k=1}^{N-1} X_1^k \right) \oplus \left(\bigoplus_{k=1}^{N-1} x_k x_{N}\right)\nonumber\\
    &= \left(\bigoplus_{k=1}^{N-1} X_1^k \right) \oplus \left(\bigoplus_{k=1}^{N-1} ( X_1^k \oplus X_2^k ) x_{N}\right).\label{multiguess}
\end{align}

\noindent Therefore, if the receiver chooses his measurement as $x_N = j$, when $j=0$ we have $G_0 = X_1^1 \oplus X_1^2 \oplus \cdots \oplus X_1^{N-1}$, and for $j=1$ we obtain $G_1 = X_2^1 \oplus X_2^2 \oplus \cdots \oplus X_2^{N-1}$. \ie,  the receiver always computes the functions perfectly and trivializes the communication task. It is clear that the task success is related to the probability of the non-signaling boxes working just as \eqref{multi45}, \ie,  $p(a_0 \oplus a_1 \oplus \cdots \oplus a_{N-1} = x_0 x_{N-1} \oplus x_1 x_{N-1} \oplus \cdots \oplus x_{N-2} x_{N-1} | x_0, x_1, \cdots, x_{N-1})$. Thus, the probabilities that the receiver computes the function values $f_1$ and $f_2$ correctly are, respectively, given by

\begin{subequations}\label{pt1pt2M}
\begin{align}
   P_I &= \frac{1}{2^{N-1}}\left[\sum_{x_1,...,x_{N-1}} p\left(\bigoplus_{k=1}^{N} a_k = \bigoplus_{k=1}^{N-1} x_k x_{N} | x_1,...,x_{N-1}, x_N =0\right)\right];\\
  P_{II} &= \frac{1}{2^{N-1}}\left[\sum_{x_1,...,x_{N-1}} p\left(\bigoplus_{k=1}^{N} a_k = \bigoplus_{k=1}^{N-1} x_k x_{N} | x_1,...,x_{N-1}, x_N =1\right)\right].
 \end{align}
 \end{subequations}
 
 \noindent When the parts share \eqref{multi45}, we have $P_I = P_{II} = 1$. However, by introducing a parameter $E\in [0,1]$, we can investigate other non-signaling behaviors by means of the following probability of success:
 \begin{align}\label{multibias}
     p\left(\bigoplus_{k=1}^{N} a_k = \bigoplus_{k=1}^{N-1} x_k x_{N}\right) = \frac{1}{2} (1 + E).
 \end{align}
The perfect correlations of behavior \eqref{multi45} are retrieved when $E=1$, and uniform probabilities are retrieved when $E=0$.
    
From this example, one can see that the concatenation approach, depicted in Fig. \ref{concatenation}, can also be employed in this multipartite scenario. This is due to the fact that, to complete the task, it is sufficient for the receiver to know only $\bigoplus_{k=1}^{N-1} M_k$, instead of each message $M_k$. For instance, when $n=4$, the senders can divide their bits into two pairs and perform the encoding just as in the previous strategy. Now, if instead of sending their respective messages, $M_k^0$ and $M_k^{1}$, the parts encode them in a third NS-box \eqref{multi45} by employing \eqref{multiguess}, the receiver is able to recover perfectly one of the functions $\bigoplus_{k=1}^{N-1} M_k^{i =0,1}$. This allows the parts to perform the same decoding one more time, resulting in perfect access by the receiver to one of the functions $f_0 = X_0^1 \oplus X_0^2 \oplus \cdots \oplus X_0^{N-1}$, $f_1 = X_1^1 \oplus X_1^2 \oplus \cdots \oplus X_1^{N-1}$, $f_2 = X_2^1 \oplus X_2^2 \oplus \cdots \oplus X_2^{N-1}$, or $f_3 = X_3^1 \oplus X_3^2 \oplus \cdots \oplus X_3^{N-1}$.

In the most general scenario, the receivers have, initially, $n = 2^K$ bits, share $n-1$ perfect copies of the non-signaling resource \eqref{multi45}, and the senders and the receiver perform the strategy just as depicted in Fig. \ref{concatenation}. Here, for each part $k$, we denote the output and input of the box $i$ of the level $l$ by $a_k^{i, l}$ and $x_k^{i, l}$, respectively. Thus, we may write the guess produced by the receiver as:
\begin{align}\label{concguess}
    G_j = \left(\bigoplus_{k=0}^{N-1} M_k\right) \oplus \left(\bigoplus_{l=0}^{K-1} a_N^{i_l,l} \right),
\end{align}
where the box $i_l$ is defined in terms of the box measured in the previous level, $i_l = 2 i_{l-1} + z_l + 1$, when $l \ge 1$. In this case, the receiver performs measurements in $K$ boxes, one in each level, among which $(N-r)$ are to $z_N^{i,l} = 0$ and $r$ to $z_N^{i,l} = 1$, where $r = z_0 + z_1 + \cdots + z_{K-1}$. Just as in the single copy scenario, the task success is directly related to the probability that the $n-1$ non-signaling boxes behave as \eqref{multi45}, \ie,  \eqref{multibias}. Thus, when $E<1$, for each box, there exists a probability that the receiver output $a_{N-1}^{i,l}$ is wrong and the property $\bigoplus_{k=1}^{N-1} a_k^{i,l} = \bigoplus_{k=1}^{N-2} x_k^{i,l} x_{N}^{i,l}$ does not hold. However, if an even number of mistakes is produced in the outputs of the receiver, then they all cancel each other and the produced guess with \eqref{concguess} will be correct. Therefore, the success probability for the multipartite task with concatenation is equal to the probability that the receiver produces an even number of wrong outputs, \ie:
\begin{align}\label{successT}
    p\left(G_j = \bigoplus_{k=0}^{N-2} X_j^k \right) = Q_{\text{even}}^{(K-r)}(P_I)\cdot Q_{\text{even}}^{(r)}(P_{II}) + Q_{\text{odd}}^{(K-r)}(P_I)\cdot Q_{\text{odd}}^{(r)}(P_{II}),
 \end{align}
where $P_I$ and $P_{II}$ are defined in \eqref{pt1pt2M} and $Q_{\text{even}}^{(s)}(P)$ and $Q_{\text{odd}}^{(s)}(P)$ are given by
\begin{subequations}\label{Q}
\begin{align}
    Q_{\text{even}}^s(P) &= \sum_{j=0}^{\lfloor\frac{s}{2} \rfloor} \binom{s}{2j} (1-P)^{2j} P^{s-2j} = \frac{1}{2}(1+(2P-1)^s);\label{Qeven}\\
    Q_{\text{odd}}^s(P) &= \sum_{j=0}^{\lfloor \frac{s-1}{2}\rfloor} \binom{s}{2j+1} (1-P)^{2j+1} P^{s-2j-1} = \frac{1}{2}(1-(2P-1)^s).\label{Qodd}
\end{align}
\end{subequations}
These describe the probabilities of the receiver producing an even and an odd number of mistakes, respectively, after $s$ measurements; $P$ denotes the probability of obtaining the right output in a NS-box. 

By inserting \eqref{Q} in \eqref{successT} and considering the bias from \eqref{multibias} in the probabilities from \eqref{pt1pt2M}, we find the communication task success probability
\begin{align}\label{concsuccess}
    p\left(G_j = \bigoplus_{k=0}^{N-2} X_j^k \right) =  \frac{1}{2}(1+E_I^{K-r}E_{II}^{r}),
\end{align}
where $E_i = 2 P_i - 1$.

\section{Proving new IC criteria}\label{truthfulness}

In this appendix, we prove criterion \eqref{ICmulti}, however, for the even more general multipartite scenario described in appendix \ref{multicommunicationtask}. The strategy will be similar to the one employed in the first bipartite proposal \cite{IC}, so, for completeness, we start by defining the following mutual information chain rule and data processing inequalities:
\begin{align}
    &I(A:B|C) = I(A:B,C) - I(A:C); \label{chain}\\
    &I(A:B') \le I(A:B), \quad \text{where} \quad B \longrightarrow B';\label{DP}
\end{align}

Following the description given in appendix \ref{multicommunicationtask}, first, we consider the following quantity, $I(\mathbf{x}^k : \mathbf{x}^1, \cdots, \mathbf{x}^{k-1}, \mathbf{x}^{k+1}, \cdots, \mathbf{x}^{N-1}, M_k, c )$, and prove that it is lower bounded by the left-hand side of \eqref{ICmulti}. By applying the chain rule \eqref{chain} two times, we obtain:
\begin{multline}
    I(\mathbf{x}^k : \mathbf{x}^1, \cdots, \mathbf{x}^{k-1}, \mathbf{x}^{k+1}, \cdots, \mathbf{x}^{N-1}, M_k, c ) \\  =  I(X_1^k : \mathbf{x}^1, \cdots, \mathbf{x}^{k-1}, \mathbf{x}^{k+1}, \cdots, \mathbf{x}^{N-1}, M_k, c )+I(X_2^k, \cdots, X_n^k : \mathbf{x}^1, \cdots, \mathbf{x}^{k-1}, \mathbf{x}^{k+1}, \cdots, \mathbf{x}^{N-1}, M_k, c | X_1^k) \\
     = I(X_1^k : \mathbf{x}^1, \cdots, \mathbf{x}^{k-1}, \mathbf{x}^{k+1}, \cdots, \mathbf{x}^{N-1}, M_k, c)+ I(X_2^k, \cdots, X_n^k : \mathbf{x}^1, \cdots, \mathbf{x}^{k-1}, \mathbf{x}^{k+1}, \cdots, \mathbf{x}^{N-1}, M_k, c, X_1^k) \\ - I(X_2^k, \cdots, X_n^k :  X_1^k).\label{two}
\end{multline}

\noindent From data processing \eqref{DP} we have 
\begin{multline}
I(X_2^k, \cdots, X_n^k : \mathbf{x}^1, \cdots, \mathbf{x}^{k-1}, \mathbf{x}^{k+1}, \cdots, \mathbf{x}^{N-1}, M_k, c, X_1^k) \ge \\ I(X_2^k, \cdots, X_n^k : \mathbf{x}^1, \cdots, \mathbf{x}^{k-1}, \mathbf{x}^{k+1}, \cdots, \mathbf{x}^{N-1}, M_k, c).
\end{multline}
Furthermore, by applying the chain rule in the first term in the right-hand side of \eqref{two}, and using strong subadditivity, $I(A:B|C) \ge 0$, we obtain: 
\begin{multline}
    I(X_1^k : \mathbf{x}^1, \cdots, \mathbf{x}^{k-1}, \mathbf{x}^{k+1}, \cdots, \mathbf{x}^{N-1}, M_k, c) \\ = I(X_1^k : X_2^1, \cdots,X_n^1| X_1^1, \mathbf{x}^{2}, \cdots, \mathbf{x}^{k-1}, \mathbf{x}^{k+1}, \cdots, \mathbf{x}^{N-1}, M_k, c) \\ + I(X_1^k : X_1^1, \mathbf{x}^{2}, \cdots, \mathbf{x}^{k-1}, \mathbf{x}^{k+1}, \cdots, \mathbf{x}^{N-1}, M_k, c) \\
    \ge I(X_1^k : X_1^1, \mathbf{x}^{2}, \cdots, \mathbf{x}^{k-1}, \mathbf{x}^{k+1}, \cdots, \mathbf{x}^{N-1}, M_k, c).\label{four}
\end{multline}
Therefore, back to \eqref{two}, we write:
\begin{multline}
    I(\mathbf{x}^k : \mathbf{x}^1, \cdots, \mathbf{x}^{k-1}, \mathbf{x}^{k+1}, \cdots, \mathbf{x}^{N-1}, M_k, c ) \\ \ge I(X_1^k : X_1^1, \mathbf{x}^{2}, \cdots, \mathbf{x}^{k-1}, \mathbf{x}^{k+1}, \cdots, \mathbf{x}^{N-1}, M_k, c) + I(X_2^k, \cdots, X_n^k : \mathbf{x}^1, \cdots, \mathbf{x}^{k-1}, \mathbf{x}^{k+1}, \cdots, \mathbf{x}^{N-1}, M_k, c) \\ - I(X_2^k, \cdots, X_n^k :  X_1^k). \label{five}
\end{multline}
Similarly to \eqref{four}, we can employ the chain rule and strong subadditivity $N-3$ times in the first right-hand side term in \eqref{five} in order to highlight only the first bit $X_1^k$ of each bit-string $\mathbf{x}^k$: 
\begin{multline}
    I(\mathbf{x}^k : \mathbf{x}^1, \cdots, \mathbf{x}^{k-1}, \mathbf{x}^{k+1}, \cdots, \mathbf{x}^{N-1}, M_k, c ) \\ \ge I(X_1^k : X_1^1, X_1^2, \cdots,X_1^{k-1}, X_1^{k+1}, \cdots, X_1^{N-1}, M_k, c)+ I(X_2^k, \cdots, X_n^k : \mathbf{x}^1, \cdots, \mathbf{x}^{k-1}, \mathbf{x}^{k+1}, \cdots, \mathbf{x}^{N-1}, M_k, c) \\- I(X_2^k, \cdots, X_n^k :  X_1^k). \label{six}
\end{multline}
Notice that the right side third term in \eqref{six} is, exactly, $I(\mathbf{x}^k : \mathbf{x}^1, \cdots, \mathbf{x}^{k-1}, \mathbf{x}^{k+1}, \cdots, \mathbf{x}^{N-1}, M_k, c )$, but without $X_1^k$ of the bit-string $x^k$. Therefore, by performing the same steps $n-1$ times, we achieve:
\begin{multline}
    I(\mathbf{x}^k : \mathbf{x}^1, \cdots, \mathbf{x}^{k-1}, \mathbf{x}^{k+1}, \cdots, \mathbf{x}^{N-1}, M_k, c ) \\ \ge \sum_i^n I(X_i^k : X_i^1, X_i^2, \cdots,X_i^{k-1}, X_i^{k+1}, \cdots, X_i^{N-1}, M_k, c) - \sum_i^n I(X_{i+1}^k, \cdots, X_n^k :  X_i^k). \label{seven}
\end{multline}
From the data processing inequality \eqref{DP}, we write 
\begin{align}
I(X_i^k : X_i^1, X_i^2, \cdots,X_i^{k-1}, X_i^{k+1}, \cdots, X_i^{N-1}, M_k, c) \ge I(X_i^k : X_i^1, X_i^2, \cdots,X_i^{k-1}, X_i^{k+1}, \cdots, X_i^{N-1}, G_i),
\end{align}
and, finally, obtain the lower bound:
\begin{multline}
    I(\mathbf{x}^k : \mathbf{x}^1, \cdots, \mathbf{x}^{k-1}, \mathbf{x}^{k+1}, \cdots, \mathbf{x}^{N-1}, M_k, c ) \\ \ge \sum_i^n I(X_i^k : X_i^1, X_i^2, \cdots,X_i^{k-1}, X_i^{k+1}, \cdots, X_i^{N-1}, G_i)- \sum_i^n I(X_{i+1}^k, \cdots, X_n^k :  X_i^k). \label{eight}
\end{multline}

The next step will be to prove that $I(\mathbf{x}^k : \mathbf{x}^1, \cdots, \mathbf{x}^{k-1}, \mathbf{x}^{k+1}, \cdots, \mathbf{x}^{N-1}, M_k, c ) \le H(M_k)$. So:
\begin{multline}
    I(\mathbf{x}^k : \mathbf{x}^1, \cdots, \mathbf{x}^{k-1}, \mathbf{x}^{k+1}, \cdots, \mathbf{x}^{N-1}, M_k, c ) \\ = I(\mathbf{x}^k : M_k | \mathbf{x}^1, \cdots, \mathbf{x}^{k-1}, \mathbf{x}^{k+1}, \cdots, \mathbf{x}^{N-1}, c)+ I(\mathbf{x}^k : \mathbf{x}^1, \cdots, \mathbf{x}^{k-1}, \mathbf{x}^{k+1}, \cdots, \mathbf{x}^{N-1}, c) \\
    = I(M_k : \mathbf{x}^1, \cdots, \mathbf{x}^{N-1}, c) - I(M_k : \mathbf{x}^1, \cdots, \mathbf{x}^{k-1}, \mathbf{x}^{k+1}, \cdots, \mathbf{x}^{N-1}, c )\\
    \le  I(M_k : \mathbf{x}^1, \cdots, \mathbf{x}^{N-1}, c),\label{nine}
\end{multline}
where here we applied the chain rule two times, considering the non-signaling between the $N$ parts and the non-negativity of the mutual information, $I(A:B)\ge 0$. At this point, just as argued in Ref.~\cite{IC}, from the data processing inequality, we have $I(M_k : \mathbf{x}^1, \cdots, \mathbf{x}^{N-1}, c) \le I(M_k : M_k) = H(M_k)$, which finally yields:
\begin{align}
    I(\mathbf{x}^k : \mathbf{x}^1, \cdots, \mathbf{x}^{k-1}, \mathbf{x}^{k+1}, \cdots, \mathbf{x}^{N-1}, M_k, c ) \le H(M_k).\label{ten}
\end{align}

Now, we can put \eqref{eight} and \eqref{ten} together in order to achieve

\begin{align}
    \sum_i^n I(X_i^k : X_i^1, \cdots,X_i^{k-1}, X_i^{k+1}, \cdots, X_i^{N-1}, G_i)  \le H(M_k) + \sum_i^n I(X_{i+1}^k, \cdots, X_n^k :  X_i^k).\label{eleven}
\end{align}

\noindent Finally, we recover \eqref{ICmulti} by summing inequality \eqref{eleven} over $k$ and considering non-signaling between the $N$ parts, \ie, $\sum\limits_k^{N-1} H(M_k) = H(M_1,...,M_{N-1})$:
\begin{align}
    \sum_k^{N-1} \sum_i^n I(X_i^k : X_i^1, \cdots,X_i^{k-1}, X_i^{k+1}, \cdots, X_i^{N-1}, G_i)  \le H(M_1,\cdots,M_{N-1})+ \sum_k^{N-1} \sum_i^n I(X_{i+1}^k, \cdots, X_n^k :  X_i^k).\label{twelve}
\end{align}

\section{Multiple copies inequality}\label{multiplebits}

Here we prove the multipartite generalization of the multiple copies criterion \eqref{triMultiple}, firstly derived in Ref.~\cite{IC} for a strict bipartite scenario. 

First of all, we need to prove a simplified lower bound for \eqref{ICmulti}. So, rewriting the left-hand side summation argument in \eqref{ICmulti}, we have 
\begin{align}
    I(X_i^k : X_i^1, \cdots,X_i^{k-1}, X_i^{k+1}, \cdots, X_i^{N-1}, G_i) = &H(X_i^k) - H(X_i^k | X_i^1, X_i^2 \cdots,X_i^{k-1}, X_i^{k+1}, \cdots, X_i^{N-1}, G_i)\nonumber\\
    = & 1 - H(X_i^k \oplus X_i^1| X_i^1, X_i^2, \cdots,X_i^{k-1}, X_i^{k+1}, \cdots, X_i^{N-1}, G_i)\nonumber \\
    \ge & 1 - H(X_i^k \oplus X_i^1| X_i^2, \cdots,X_i^{k-1}, X_i^{k+1}, \cdots, X_i^{N-1}, G_i).\label{thirteen}
\end{align}
Here, we particularized to the case where every bit $X_i^k$ is associated with a uniform distribution, $H(X_i^k) = 1$. Further, we considered the fact that $H(A|B, C) = H(A \oplus B | B, C)$, because knowing $B$ results in the same uncertainty about $A$ and $A\oplus B$, and $H(A\oplus B | B, C) \ge H(A\oplus B | C)$, \ie, to remove the conditioning in $B$ does not increase the uncertainty of $A \oplus B$. This same argument can be applied $N-2$ times in order to move every conditioned random variable in the right-hand side of \eqref{thirteen}:
\begin{align}
    I(X_i^k : X_i^1, \cdots,X_i^{k-1}, X_i^{k+1}, \cdots, X_i^{N-1}, G_i) \ge  1 - H(X_i^1 \oplus X_i^2 \oplus \cdots \oplus  X_i^{N-1} \oplus G_i).\label{fourteen}
\end{align}
However, from the communication task, when $X_i^1 \oplus X_i^2 \oplus \cdots \oplus  X_i^{N-1} \oplus G_i = 0$, we necessarily have $G_i = X_i^1 \oplus X_i^2 \oplus \cdots \oplus  X_i^{N-1}$. Thus, the probability $p(X_i^1 \oplus X_i^2 \oplus \cdots \oplus  X_i^{N-1} \oplus G_i = 0)$ is exactly the success probability of the receiver, $p(G_i = X_i^1 \oplus X_i^2 \oplus \cdots \oplus  X_i^{N-1})$, while $p(X_i^1 \oplus X_i^2 \oplus \cdots \oplus  X_i^{N-1} \oplus G_i = 1)$ is the complementary part. Therefore, the right-hand side term from \eqref{fourteen} can be written in terms of the binary entropy, which in \eqref{twelve} finally yields: 
\begin{align}
    (N-1) \sum_i^n (1 - h(p(G_i = X_i^1 \oplus X_i^2 \oplus \cdots \oplus  X_i^{N-1}))) \le \mathcal{I} \le H(M_1,\cdots,M_{N-1}). \label{fifteen}
\end{align}
Notice that we considered the fact that the left-hand side has no dependence on the index $k$. Furthermore, the rightmost term in \eqref{twelve} does not appear in \eqref{fifteen}, because we are assuming a uniform distribution for every initial bit $X_i^k$.

At this point, we particularize our description to the concatenation strategy earlier described in appendix \ref{multicommunicationtask}. Here we rewrite the left-hand side summation in \eqref{fifteen} in terms of the number of instances $r$ where the receiver performed measurement ${x_n}_j^k = 1$, and substitute the concatenation success probability \eqref{concsuccess}:
\begin{align}
    (N-1) \sum_i^n (1 - h(p(G_i = X_i^1 \oplus X_i^2 \oplus \cdots \oplus  X_i^{N-1}))) &= (N-1) \sum_r^K \binom{K}{r} \left[ 1 - h\left(\frac{1+E_I^{K-r} E_{II}^{r}}{2}\right)\right]\nonumber\\
    &\ge \frac{(N-1)}{2\ln 2} \sum_r^K \binom{K}{r} (E_I^2)^{N-r}(E_{II}^2)^r \nonumber \\
    & = \frac{(N-1)}{2\ln 2} (E_I^2 + E_{II}^2)^K
    \label{sixteen},
\end{align}
where we considered $1-h\left(\frac{1+y}{2}\right) \ge \frac{y^2}{2\ln 2}$ and $E_i = 2P_i -1$, from \eqref{pt1pt2M}. After performing such encoding, each sender sends only a single bit message. Thus, $H(M_1,\cdots,M_{N-1}) $ in \eqref{fifteen} is always fixed in $N-1$, necessarily. Therefore, with \eqref{fifteen} and \eqref{sixteen}, we find that when $E_I^2 + E_{II}^2 > 1$, the new proposed criterion \eqref{ICmulti} can always be violated by some concatenation protocol with $K$ levels. Thus, we finally conclude the proof for the previously mentioned criterion in \eqref{triMultiple}:
\begin{align}\label{seventeen}
    E_I^2 + E_{II}^2 \le 1.
\end{align}

\section{New inequality in terms of noisy channel capacity}\label{noisyinequality}

Here we prove the inequality \eqref{ICmultiNoisy}, where the senders communicate their messages $M_k$ through a single use of a noisy channel to the receiver. The proof for the noiseless version \eqref{ICmulti} is essentially valid in this context, but it is necessary to introduce a new variable $M_k '$, representing the message after the action of the channel, on step \eqref{nine} to obtain the upper bound in terms of the channel capacity $C_k = I(M_k : M_k ')$.  So, we have:
\begin{align}
    I(\mathbf{x}^k : \mathbf{x}^1, \cdots, \mathbf{x}^{k-1}, \mathbf{x}^{k+1}, \cdots, \mathbf{x}^{N-1}, M_{k}', c ) 
    \le & I(M_{k}' : \mathbf{x}^1, \cdots, \mathbf{x}^{N-1}, c)\nonumber\\
    \le & I(M_{k}' : \mathbf{x}^1, \cdots, \mathbf{x}^{N-1}, c, M_{k}),\label{eighteen}
\end{align}
where we considered the data processing inequality in the second step. As mentioned in the main text, $M_k$ completely determines $M_k '$, so $M_k '$ is conditionally independent of any random variable from the causal structure, \ie, $\;I(M_{k}' : V | M_{k}) = 0$. Thus, we may write $I(M_{k}' : \mathbf{x}^1, \cdots, \mathbf{x}^{N-1}, c, M_{k}) - I(M_{k}' : M_{k}) = 0$ and obtain in \eqref{eighteen}
\begin{align}
    I(\mathbf{x}^k : \mathbf{x}^1, \cdots, \mathbf{x}^{k-1}, \mathbf{x}^{k+1}, \cdots, \mathbf{x}^{N-1}, M_{k}', c ) 
    \le & I(M_{k}' : M_{k})  = C_k. \label{nineteen}
\end{align}
The next steps are quite similar as the appendix \ref{truthfulness}, therefore we may write
\begin{align}
   \sum_k^{N-1} \sum_i^n I(X_i^k : X_i^1, \dots,X_i^{k-1}, X_i^{k+1}, \dots, X_i^{N-1}, G_i) \le \sum_k^{N-1} C_k + \sum_i^n I(X_{i+1}^k, \dots, X_n^k :  X_i^k).
\end{align}

\end{document}